\documentclass[aps,pra,amsmath,amssymb,amsfonts,superscriptaddress,lengthcheck,twocolumn,longbibliography]{revtex4-1}
\usepackage{dcolumn}    
\usepackage{graphicx}
\usepackage{amsmath, bbm}
\usepackage{latexsym}
\usepackage{amsfonts}   
\usepackage{amssymb}
\usepackage{array}      
\usepackage{epsfig}
\usepackage{txfonts}
\usepackage{xcolor,braket}
\usepackage[colorlinks=true,linkcolor=blue,urlcolor=blue,citecolor=blue,pdfusetitle]{hyperref}
\usepackage{hyperref}
\usepackage{ulem}
\usepackage{soul}
\usepackage{dsfont}

\usepackage{relsize}

\usepackage{mathrsfs}

\usepackage{cancel}
\usepackage{enumitem}

\newcommand{\dyad}[1]{\left\vert#1\right\rangle\!\left\langle#1\right\vert}
\newcommand{\tr}[1]{\mathrm{tr}\left\{#1\right\}}
\newcommand{\steve}[1]{{\color{black}#1}}

\newcommand{\eoin}[1]{{\color{black}#1}}

\begin{document}
 \title{Stochastic collisional quantum thermometry} 
       
\author{Eoin O'Connor}
\affiliation{School of Physics, University College Dublin, Belfield, Dublin 4, Ireland}
\affiliation{Centre for Quantum Engineering, Science, and Technology, University College Dublin, Belfield, Dublin 4, Ireland}
\author{Bassano Vacchini}
\affiliation{Dipartimento di Fisica ``Aldo Pontremoli", Universit{\`a} degli Studi di Milano, via Celoria 16, 20133 Milan, Italy}
\affiliation{Istituto Nazionale di Fisica Nucleare, Sezione di Milano, Via Celoria 16, 20133 Milan, Italy}
\author{Steve Campbell}
\affiliation{School of Physics, University College Dublin, Belfield, Dublin 4, Ireland}
\affiliation{Centre for Quantum Engineering, Science, and Technology, University College Dublin, Belfield, Dublin 4, Ireland}
\begin{abstract}
We extend collisional quantum thermometry schemes to allow for stochasticity in the waiting time between successive collisions. We establish that introducing randomness through a suitable waiting time distribution, the Weibull distribution, allows to significantly extend the parameter range for which an advantage over the thermal Fisher information is attained. These results are explicitly demonstrated for dephasing interactions and also hold for partial swap interactions. Furthermore, \steve{we show that the optimal measurements can be performed locally}, thus implying that genuine quantum correlations do not play a role in achieving this advantage. We explicitly confirm this by examining the correlation properties for the deterministic collisional model. 
\end{abstract}
\date{\today}

\maketitle

\section{Introduction}
Accurately determining the temperature of a physical system is a ubiquitous task. For quantum systems, measuring the temperature becomes a significantly more involved job, in part due to the inherent fragility of quantum states, and more pointedly, due to the fact that temperature itself is not a quantum observable. Recently, significant advances in thermometry schemes for quantum systems have been proposed~\cite{Giovannetti2006, Acin2018, Braun2018, Razavian2019, Correa2015, Mitchison2020} (see Ref.~\cite{Mehboudi2019} for an extensive review). The thermometric precision of a probe in equilibrium with the sample is limited by the thermal Cramer-Rao bound, which is inversely proportional to the heat capacity, $C$, of the thermometer: $(\Delta T/T)^2 \geq k_B/NC$. However, quantum systems have the additional freedom to exploit resources such as entanglement~\cite{Maccone2013, Huang2016} and coherence~\cite{Micadei2015, Pires2018, Castellini2019} to gain an advantage over their classical counterpart~\cite{Campbell2017, Jahromi2020, Correa2017, Hovhannisyan2018, Ivanov2019, Jevtic2015, Kiilerich2018, Mancino2017, Potts2019}. By making use of these resources, along with collective measurements on multiple probes, it is possible to surpass the $1/N$ scaling of the Cramer-Rao bound.

A typical `direct' thermometry scheme will involve a number of probes coupled to the system of interest (environment) and, after a suitable interaction, these probes are measured in order to estimate the temperature. However, recently an alternative approach was proposed that exploits collision models~\cite{Campbell2021a,ciccarello2021quantum, Lorenzo2017, DeChiara2020, Taranto2020, Campbell2018, McCloskey2014, DeChiara2018, Grimmer2016, Scarani2002, Strasberg2017} which involve a stream of auxiliary systems (often termed ``ancillae") interacting with an intermediary system, which is directly coupled to the environment~\cite{Seah2019}. The collisions occur for a sufficiently small time that the auxiliary systems never fully thermalise with the environment-intermediary system compound. However, information about the temperature of the environment is indirectly imprinted onto the auxiliaries and thus this scheme allows us to make use of this additional out-of-equilibrium information to enhance the precision of temperature estimation~\cite{Seah2019, Shu2020, alves2021bayesian}. 

In this paper we extend this collisional approach to allow for stochasticity in the waiting times between the collisions. We show that introducing this stocasticity leads to a broadening of range of parameters in which a meaningful advantage is gained over a direct thermometry scheme where $N$-probes thermalise with the environment and are then measured. Interestingly, \steve{for a dephasing interaction between the intermediary system and auxiliaries} we establish that the correlations generated in the collisional thermometry approach are purely classical and, therefore, the demonstrated advantage can be gained by performing simple local measurements.

The remainder of the paper is organized as follows. Sec.~\ref{sec:Fra} gives an outline of the various topics and techniques employed in collisional quantum thermometry. These techniques are then examined more closely to determine the exact role of correlations and the free parameters. In Sec.~\ref{sec:Sto} we introduce stochasticity at the level of the waiting time between collisions. We analyse how this stochasticity affects the precision of the measurements and the form of the optimal measurements. We also discuss how our results extend to different forms of measurements. Finally, our conclusions and some further discussions are presented in Sec.~\ref{sec:Concl}.
 
\section{Quantum Thermometry}
\label{sec:Fra}
\subsection{Thermal Fisher Information}
The maximum precision with which the temperature of the environment can be measured is determined by the quantum Cramer-Rao bound
\begin{equation}
    (\Delta T)^2 \geq \frac{1}{\mathcal{F}(T,\rho)},
\end{equation}
where $\mathcal{F}(T,\rho)$ is the quantum Fisher information (QFI) of the state, $\rho$, at temperature, $T$. In order to gain information from $\rho$, a measurement, or positive operator valued measure (POVM), $\Pi$,  must be performed on the state. The outcome of this measurement is then determined by the probability distribution $p(x) \!=\! \text{tr} (\Pi_x \rho$) where $\Pi_x$ is the POVM element associated with measurement outcome $x$. The Fisher information associated with this measurement is given by
\begin{equation}
    F(T,\Pi,\rho) = \sum_x p(x)\left(\frac{\partial}{\partial T}\ln p(x)\right)^2.
\end{equation}
The QFI is attained by maximising this Fisher information over all POVMs. This maximisation can be determined from the expectation value of the square of the symmetric logarithmic derivative, $\mathcal{F}(T,\rho) \!=\! \tr{\rho\Lambda^2}$, with $\Lambda$ defined implicitly by $ 2 \partial_t \rho \!=\! \Lambda \rho + \rho \Lambda$. 

For a typical thermometry scheme involving a number of probes fully thermalising with the environment; the corresponding QFI is known as the thermal Fisher information and is given by
\begin{equation}
    \mathcal{F}(T,\rho) = \mathcal{F}_{th} = \frac{C}{k_B T^2}, \quad\quad C = \frac{\langle H_p^2\rangle-\langle H_p\rangle^2}{k_B T^2}
\end{equation}
where $C$ is the heat capacity of the probe, $H_p$ is the probe Hamiltonian, and the probes are assumed to have reached the Gibbs state with inverse temperature, \steve{$\beta\!=\!1/k_B T$}. In the case of a qubit probe with frequency $\Omega$, the thermal Fisher information is
\begin{equation}
\label{thermalqubitFI}
    \mathcal{F}_{th} = 
    \frac{1}{\bar{n}(\bar{n}+1)(2\bar{n}+1)}\left(\frac{\partial \bar{n}}{\partial T}\right)^2
\end{equation}
where 
\begin{equation}
\bar{n} \!=\! 1/(e^{\hbar \Omega / k_B T} -1),
\end{equation}
is the mean occupation number at frequency $\Omega$ and temperature $T$. Thus, estimating $\bar{n}$ is equivalent to estimating temperature, $T$. Eq.~\eqref{thermalqubitFI} provides a lower bound which we can benchmark the performance of our stochastic collision scheme against.

\subsection{Collisional Thermometry}
Here we recap the basic ingredients of the collisional thermometry scheme outlined in Refs.~\cite{Seah2019,Shu2020}. Our set up consists of a (large) environment, $E$, at fixed temperature $T$, and it is this temperature that we wish to estimate. The environment is coupled to an intermediary system, $S$, such that in the absence of any other interaction, $S$ will reach thermal equilibrium with $E$. This intermediary system is in turn coupled to a stream of independent and identically prepared auxiliary units, $A_i$ which form the collisional bath. In what follows, we will assume both $S$ and all $A_i$'s are qubits and that the $S$-$A_i$ interaction is unitary. Information about the temperature is then gained by performing measurements on the $A_i$'s, either individually or in batches. This is in contrast to standard probe based thermometry, where the probes interact directly with the environment, with the best precision occurring when they are permitted to thermalise fully before being measured. We assume that the $S$-$A_i$ interaction time, $\tau_{SA}$, is small compared to the system-environment coupling time, $\tau_{SE}$ allowing us to neglect the system-environment coupling during the collisions. After $N$ collisions the system and auxiliaries are given by the combined state
\begin{equation}
    \rho_{S,A_1,...,A_N} = \mathscr{U}_{S A_N}\circ\steve{\mathcal{E}\circ \mathscr{U}_{S A_{N-1}}\circ}...\circ\mathcal{E}\circ\mathscr{U}_{S A_1}(\rho_S \otimes \rho_{A_1}  \otimes ...  \otimes \rho_{A_N})
    \label{evo}
\end{equation}
where $\mathscr{U}_{S A_i}(\circ) \!=\! {U}_{S A_i}\circ {U}_{S A_i}^\dagger$ \steve{and $\mathcal{E}$ corresponds to the map induced by the $S$-$E$ interaction acting on the intermediary system in between the collisions.}

We model the $S$-$E$ interaction by a general thermalising master equation in the weak coupling limit, which in the interaction picture is given by
\begin{equation}
    \frac{d \rho_S}{dt} = \mathcal{L(\rho_S)} =\gamma(\bar{n}+1)\mathcal{D}[\sigma_-^S](\rho_S)+\gamma \bar{n}\mathcal{D}[\sigma_+^S](\rho_S)
    \label{master}
\end{equation}
where $\mathcal{D}[A](\rho) \!=\! A\rho A^\dagger-\frac12 \{A^\dagger A,\rho\}$, $\gamma$ is the system-environment coupling constant. We can now calculate $\mathcal{E}$ \steve{in} Eq.~\eqref{evo} by integrating Eq.~\eqref{master} over the time between subsequent collisions $\tau_{SE}$. The resulting channel takes the form $\mathcal{E} \! = \! e^{\tau_{SE}\mathcal{L}}$. This is a thermalising map which brings $S$ towards the Gibbs state, $\rho_S^{th}$, i.e. $\mathcal{E}(\rho_S^{th})\! =\! \rho_S^{th}$. We choose the intermediary system and auxiliaries to be resonant, i.e. $H_S \!=\! H_A \!=\! \hbar \Omega \sigma_z/2$. The system, $S$, therefore experiences the stroboscopic map
\begin{equation}
    \rho_S^i = \text{tr}_{A_i}\{\mathscr{U}_{S A_i}\circ\mathcal{E}(\rho_S^{i-1}\otimes\rho_{A_i})\} := \Phi(\rho_S^{i-1}).
\end{equation}
For equally spaced collision times, this map has a unique steady state $\rho_S^* \!=\! \Phi(\rho_S^*)$, which is not necessarily the Gibbs state, with the notable exception of a pure dephasing interaction between $S$ and $A_i$ as outlined in the following section.

\begin{figure*}[t]
\begin{center}
{\bf (a)}\hskip0.6\columnwidth{\bf (b)}\hskip0.6\columnwidth{\bf (c)}
    \includegraphics[width=0.65\columnwidth]{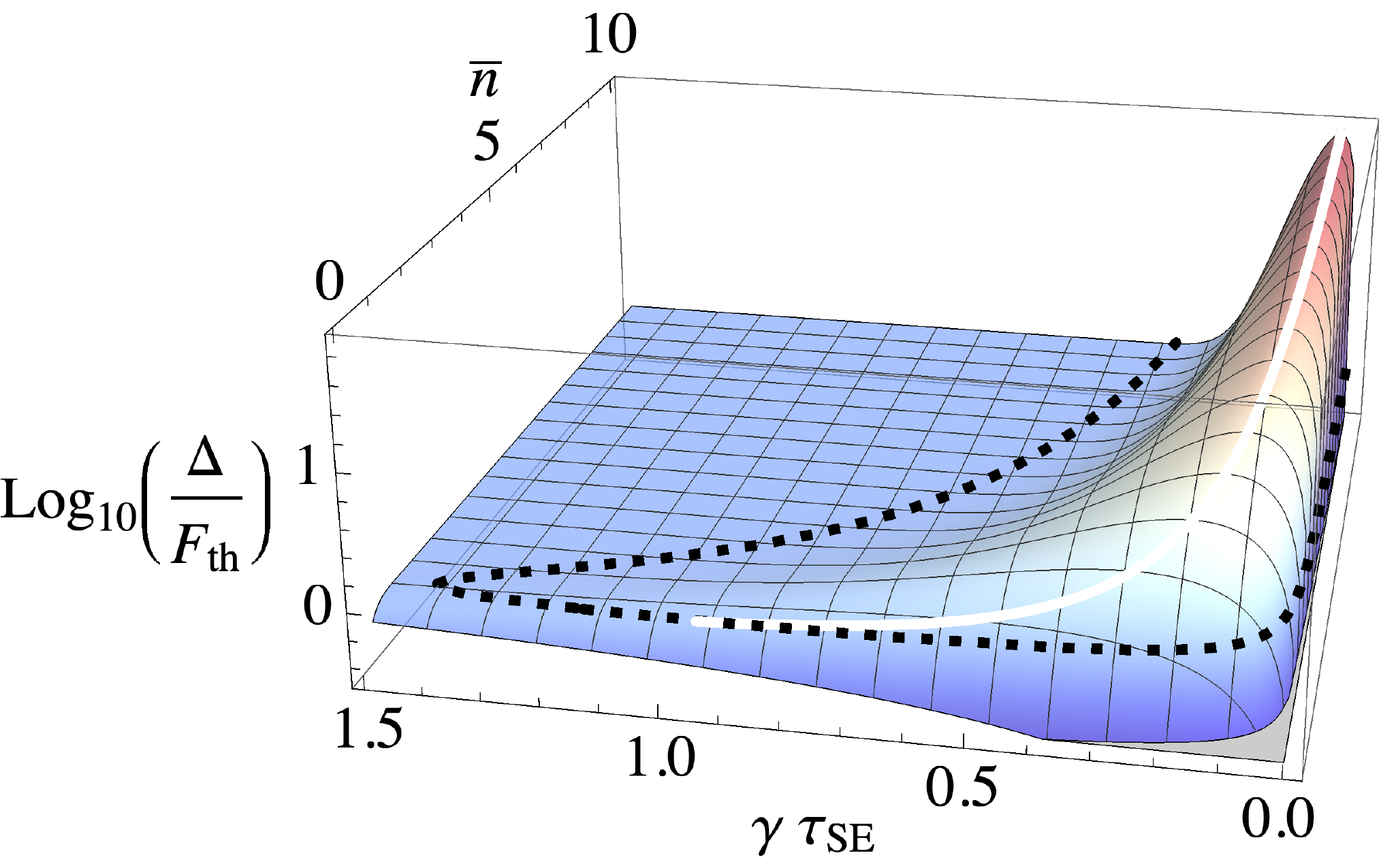}~~\includegraphics[width=0.65\columnwidth]{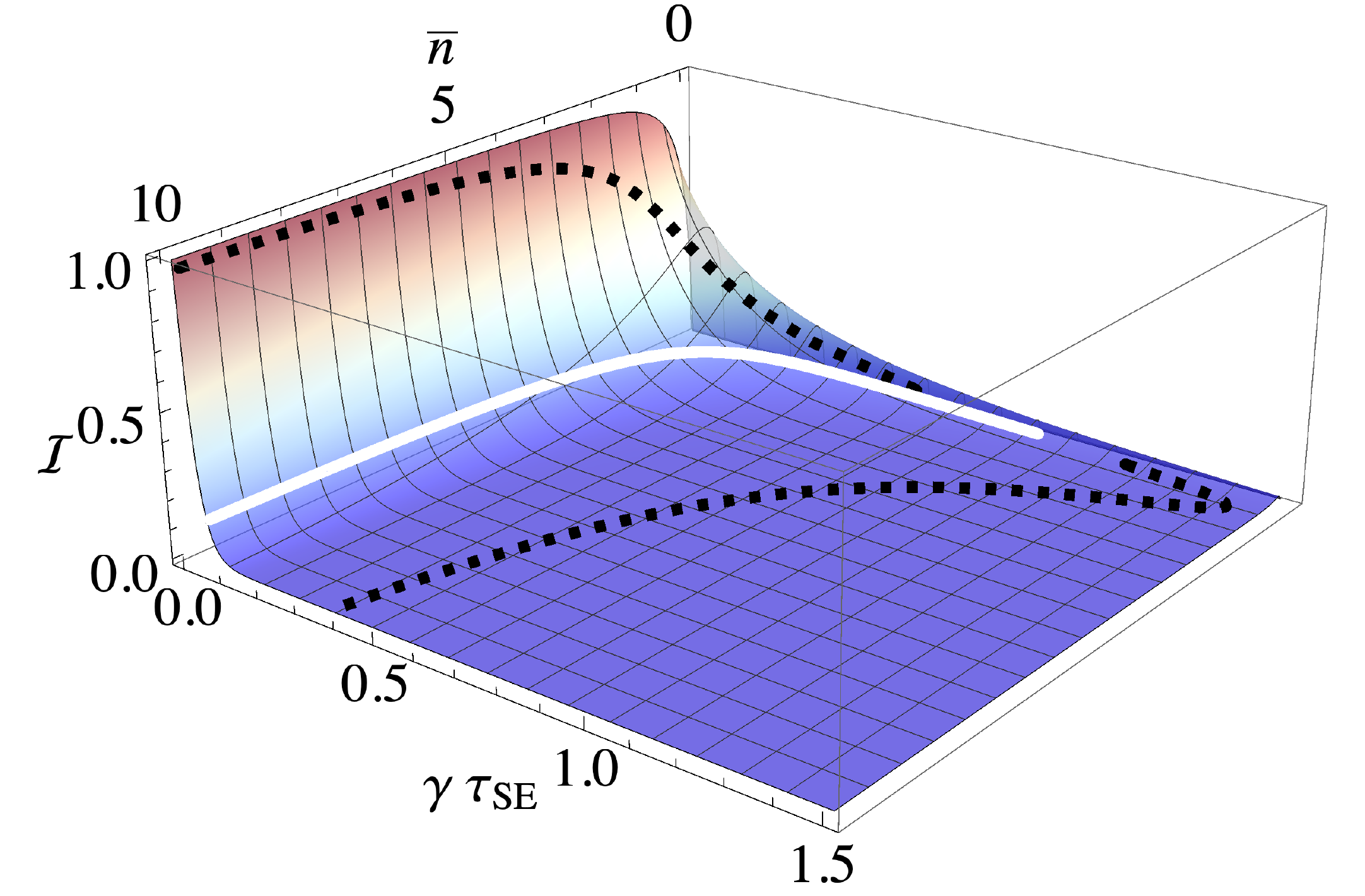}~  \includegraphics[width=0.65\columnwidth]{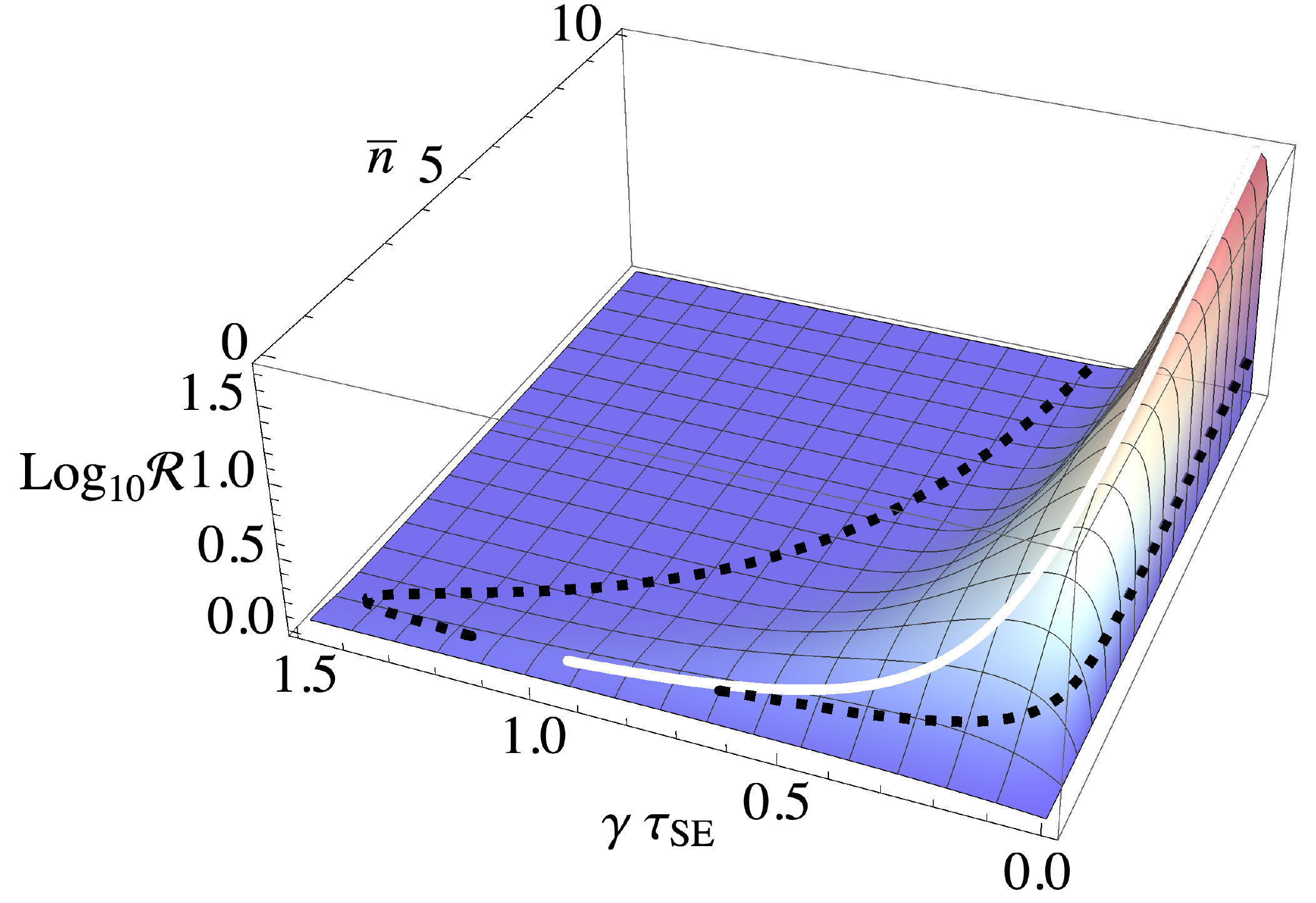}
\end{center}
\caption{{\bf (a)} Plot of the (log of the) ratio between $\Delta$, Eq.~\eqref{Eq::Delta}, and $\mathcal{F}_{th}$. Positive regions indicate parameter regimes where a thermometic advantage is achievable via the collisional scheme.
 {\bf (b)} Mutual information between two adjacent auxiliary units after each has interacted with the system via a $ZZ$ interaction for a deterministic collisional therometry protocol. {\bf (c)} Measure of the interdependence between $\gamma$ and \steve{$\Bar{n}$} captured by Eq.~\eqref{cor} for a deterministic protocol with $N\!=\!2$ (arbitrary choice). In all panels, the area captured by the dashed black line represents the region in parameter space where the scheme achieves an advantage over the thermal QFI. The white line corresponds to the value of \steve{$\Bar{n}$} where the QFI is maximal.} 
\label{fig:Determin}
\end{figure*}
\subsection{Dephasing Interactions}
We begin by focusing on a $ZZ$ interaction between the collisional bath and the system,
\begin{equation}
\label{dephasing}
H_{S A_i}^{ZZ}  = \frac{\hbar g}{2} \sigma_S^Z\sigma_{A_i}^Z,
\end{equation}
which leads to a dephasing in the energy eigenbasis and is also referred to as an indirect measurement interaction. We tune the effective system-environment coupling, $\gamma \tau_{SE}$, and the effective $S$-$A_i$ coupling, $g \tau_{SA}$. Due to the fact that Eq.~\eqref{dephasing} only affects the off-diagonal elements (i.e. coherences), and assuming that $S$ begins in thermal equilibrium with $E$ before any interactions with the auxiliaries occur, the reduced state of $S$ will remain in the Gibbs state. However, provided a suitable choice of initial state for the auxiliaries is chosen, information about the environment temperature can be imparted to the collisional bath. For a single auxiliary unit the QFI is maximised when its initial state is perpendicular to the $Z$-axis, e.g. $\ket{+_x} \!=\! (\ket{g}+\ket{e})/\sqrt{2}$ with the corresponding QFI given by
\begin{equation}
\label{thermalQFI1}
    \mathcal{F}^{\ket{+_x}} = \frac{1-\cos(2 g \tau_{S A})}{2}\mathcal{F}^{th},
\end{equation}
which is clearly maximised when the coupling parameter $g \tau_{S A} \!=\! \pi/2$. However, from Eq.~\eqref{thermalQFI1} we clearly see that \steve{for the considered interaction} it is impossible to beat the thermal Cramer-Rao bound by measuring a single auxiliary unit. When multiple $A_i$'s collide with the system in succession, correlations can be established between them, however, for long times between successive collisions, the (re)thermalisation due to system-environment interaction will destroy all correlations leaving the total QFI equal to the sum of all the individual QFIs for each $A_i$, i.e. $N$ times Eq.~\eqref{thermalQFI1}. Conversely, in the short inter-collision time limit \steve{the first collision provides as much information about the temperature as possible and, while classical correlations are established between successive $A_i$'s, no additional information can be gained about the temperature of the environment by measuring multiple auxiliary units}. Between these two extremes, relevant information about the temperature of the environment can be encoded into the auxiliary systems and provide significant advantages in precision over the thermal Cramer-Rao bound.

Fixing the optimal $S$-$A_i$ collision such that $g \tau_{S A} \!=\! \pi/2$, the QFI for $N$ auxiliaries interacting with $S$ is given by~\cite{Shu2020}
\begin{align}
    &\mathcal{F}^{\ket{+_x}}_N = \mathcal{F}^{th} + (N-1) \Delta,\qquad\text{with}, \label{QFIdeter}\\
    \Delta &= \frac{(1+\bar{n})^2\left[1-e^{\Gamma}(1+2\bar{n}\Gamma)\right]}{(1-e^{\Gamma})\left(1-\frac{(1-e^{\Gamma})\bar{n}}{1+2\bar{n}}\right)}+\frac{\bar{n}^2\left[-1+e^{\Gamma}(1-2(1+\bar{n})\Gamma)\right]}{(1-e^{\Gamma})\left(1-\frac{(1-e^{\Gamma})(1+\bar{n})}{1+2\bar{n}}\right)}
    \label{Eq::Delta}
\end{align}
where $\Gamma \!=\! \gamma(2\bar{n}+1)\tau_{SE}$ is the effective thermalisation rate of the system. From Eq.~\eqref{QFIdeter} we find that the condition for beating the thermal Cramer-Rao bound corresponds to $\Delta/\mathcal{F}_{th} \!>\! 1$, shown in Fig.~\ref{fig:Determin}(a)~\cite{Shu2020}. Furthermore, the expression for $\Delta$ demonstrates that knowledge of the $S$-$E$ coupling parameter, $\gamma \tau_{S E}$, is essential to achieve any boost in thermometric performance. For the remainder of this section we will assume a deterministic collisional scheme, namely the system and the environment interaction time is identical between each of the collisional events, i.e. $\gamma\tau_{SE}$ is the same between each collision. Thus, we consider the same setting as Refs.~\cite{Seah2019, Shu2020} of equally distributed collisions and in Sec.~\ref{sec:Sto} we introduce stochasticity.

\subsection{Role of correlations}
\label{correlations}
Given that, for the $ZZ$ interaction, measuring a single auxiliary cannot outperform the thermal Cramer-Rao bound it is natural to ask what allows for the enhancement when multiple units are measured and how this relates to the correlations established between successive $A_i$'s and/or between $S$ and a given auxiliary. We can quantify these correlations via the bipartite mutual information
\begin{equation}
\mathcal{I} = S(\rho_A) + S(\rho_B) - S(\rho_{AB}), 
\end{equation}
where $S(\cdot)$ is the von Neumann entropy. This quantity captures all correlations, both quantum and classical, present in the state. In Fig.~\ref{fig:Determin}(b) we show the mutual information shared between two successive auxiliaries, i.e. $\rho_{{A_i A_{i+1}}}$ where it clearly appears that significant correlations are established which depend on the time between each collision and the temperature of the environment. The dashed black line encloses the area in which an advantage of $>1\%$ can be gained from this setup and indicates that while there appears to be a qualitative relationship between the magnitude of the mutual information shared between the auxiliaries and the corresponding thermometric performance, with some amount of mutual information clearly being necessary in order to gain an advantage, remarkably too much correlation actually results in the QFI being lower than the thermal Fisher information. The boundary is delineated by the white line which tracks the peak QFI for each value of $\Bar{n}$. We can further characterise the type of correlations present by determining the quantum discord~\cite{Ollivier2001a,Henderson2001a} which captures the genuine quantum nature of the correlations present, and in this case turns out to be identically zero. This implies that the correlations contributing to the increased metrological performance are purely classical.


\subsection{Parameter Dependence}
\label{paradep}
We see from Eq.~\eqref{Eq::Delta} that the advantage gained from this collisional approach depends on the effective thermal relaxation parameter $\Gamma \!=\! \gamma(2\steve{\Bar{n}}+1)\tau_{SE}$. Consequently, in order to maximise this advantage {\it both} $\gamma$ and $\tau_{SE}$ must be known with certainty. While $\tau_{SE}$ corresponds to the time between collisions, which given sufficient control over the collisional bath can in principle be known, the parameter $\gamma$ which corresponds to the coupling strength between the system and the environment is more delicate. In certain circumstances it may be that, prior to any measurements, $\gamma$ is known for the setup. However, when this is not the case, or if the bath is prone to some other disturbance, determining it precisely is essential~\cite{luiz2021machine}.

We can demonstrate the importance of knowing $\gamma$ through the total variance of a measurement in multi-parameter estimation by summing the variances of all parameters. \eoin{When estimating $m$ unknown parameters we get the following chain of inequalities}~\cite{Liu2019}
\begin{equation}
    \sum_a^m \text{var}(x_a)\geq\frac 1m\tr{\mathcal{F}^{-1}}\geq \sum_a\frac{1}{m\mathcal{F}_{aa}},
\end{equation}
where $\mathcal{F}$ is the QFI matrix. \steve{In our case $m\!=\!2$ and $x_a \!\in\! \{\Bar{n}, \gamma\}$}. The second inequality is only saturated when all of the parameters are independent of each other and therefore by comparing the ratio between the second and third terms, which we denote as 
\begin{equation}
\steve{\mathcal{R} = \frac{\tr{\mathcal{F}^{-1}}}{\sum_a\frac{1}{\mathcal{F}_{aa}}},}
\label{cor}
\end{equation}
allows us to identify the areas in which knowledge of $\gamma$ is necessary in order to estimate the temperature. Figure~\ref{fig:Determin}(c) shows the peaks of this ratio line up perfectly with the peaks of the QFI. Additionally if one has no knowledge of $\gamma$ it is impossible to gain any advantage over the thermal Fisher information. In fact, the QFI is smaller than the thermal fisher information in this case when the time between collisions is small.

\section{Stochastic Approach}
\label{sec:Sto}
The previous section outlined the basic ingredients of the collisional thermometry scheme for the deterministic case introduced in Refs.~\cite{Seah2019,Shu2020}. We now turn our attention to our main focus, introducing stochasticity at the level of the time between collisions, $\tau_{SE}$, while keeping the average collision time, $tau_{SA}$, consistent with the previous section. 

\subsection{Random Collision Times}
In nature, interactions will not generally occur in fixed intervals or at deterministic times. Rather, processes are typically random with the time between interactions captured by a suitable probability distribution, the waiting time distribution (WTD). Within the framework of open quantum systems, collision models allow us to introduce such randomness either in the intervals between successive collisions or in the collision time itself, and are referred to as stochastic collision models~\cite{Strasberg2019,Vacchini2020,Chisholm2021}.

\begin{figure}[t]
\begin{center}
    \includegraphics[width=0.9\columnwidth]{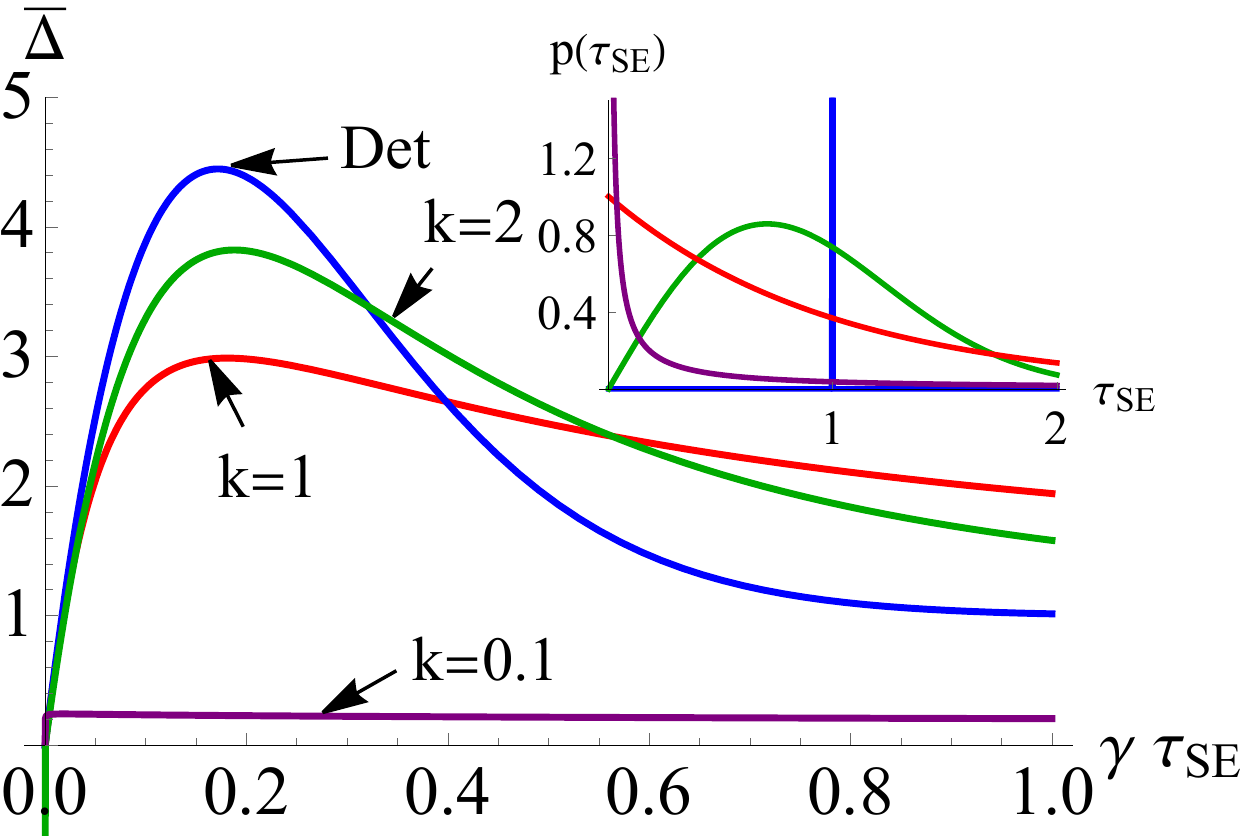}
\end{center}
\caption{Comparison of the value of the quantum Fisher information for various Weibull distributions of the collision time interval (see Eq.~\eqref{weibull}), with the deterministic case, for $\Bar{n} \!=\! 2$. Similar behavior is seen for other values of temperature above $\Bar{n}\! =\! 1.5$. \steve{{\it Inset:} Distributions for various values of $k$ shown in the main panel.}}
\label{fig:weibull}
\end{figure}

While employing WTDs in this sense clearly brings collision models closer to modelling real physical systems~\cite{Tabanera2021, Ehrich2020,Jacob2021}, here we explore how introducing such randomness affects the performance of the collisional thermometry scheme. As we shall demonstrate, stochastic collision models allow to achieve a greater range of parameter estimation over deterministic collision models without significantly sacrificing the maximal achievable precision. For concreteness, we shall focus on the Weibull renewal distribution~\footnote{We remark that our results remain qualitatively unaffected for other families of WTD, e.g. Erlang distributions.}
\begin{equation}
\label{weibull}
p(t) = \frac{k}{\lambda}\left(\frac{t}{\lambda}\right)^{k-1}e^{-(t/\lambda)^{k}},
\end{equation}
where $\lambda$ is the average time between collisions and $k$ determines the shape of the distribution, \steve{cfr. the inset of Fig.~\ref{fig:weibull}}. In particular, large $k$ tends to regular intervals between collisions, i.e. $k \!\to\! \infty$ corresponds to deterministic, equally-spaced collisions as considered in Refs.~\cite{Seah2019, Shu2020} and Secs.~\ref{correlations} and \ref{paradep}, whereas small $k$ is characterised by bursts of collisions followed by long breaks~\cite{Chisholm2021}. For $k\!=\! 1$, the WTD corresponds to the exponential distribution characterising a Poisson point process.

\begin{figure}[t]
\begin{center}
    \includegraphics[width=0.9\columnwidth]{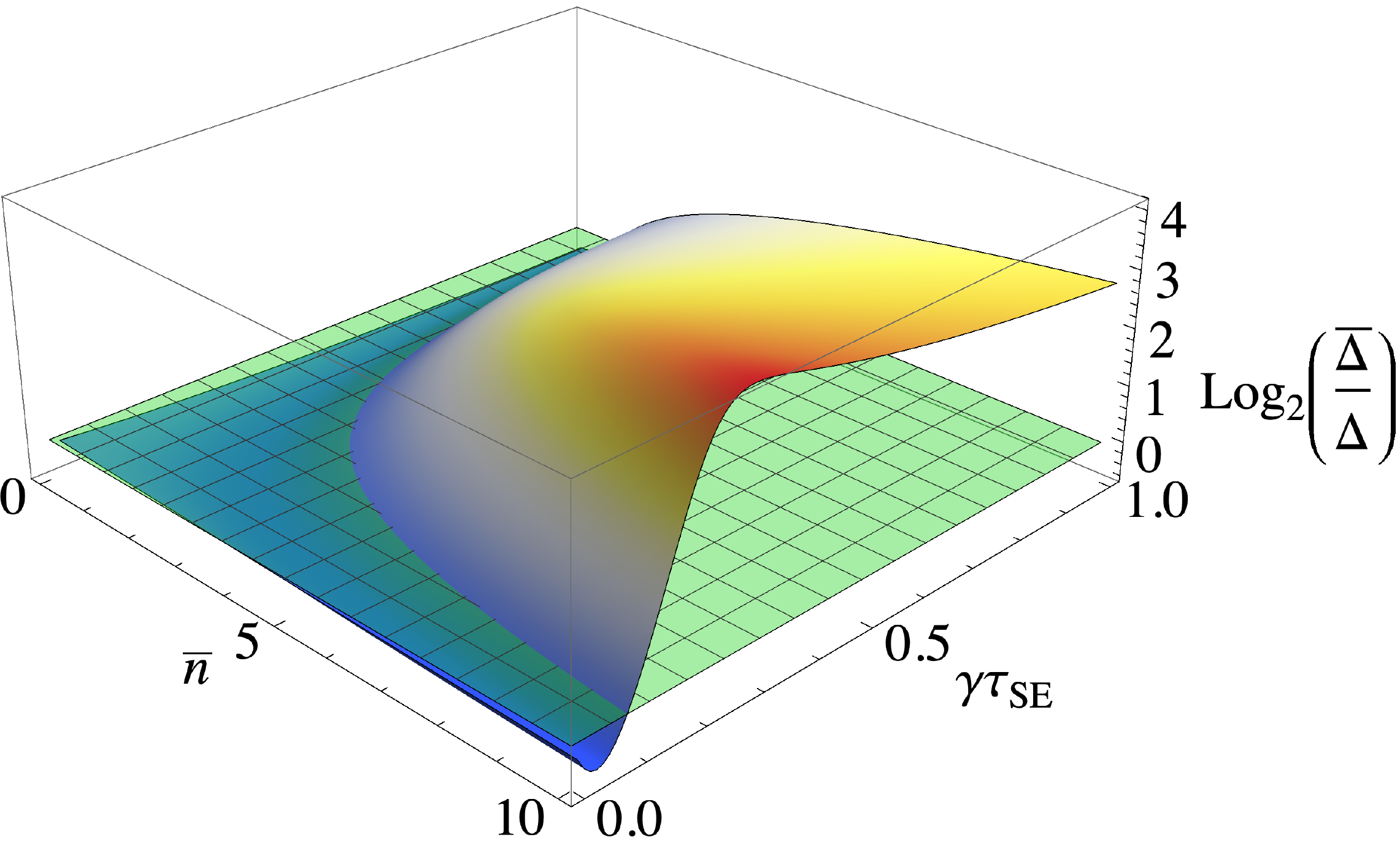}
\end{center}
\caption{Comparison of the ratio between $\Delta$ for a Weibull distribution \steve{for the exponential distribution}, i.e. $k\!=\!1$, and a deterministic equally spaced waiting time distribution. The green plane represents the crossing point where one term becomes larger than the other. $\gamma \tau_{SE}$ is the average time between collisions}
\label{fig:ZZratio}
\end{figure}

As we see from Eq.~\eqref{QFIdeter}, when the waiting time between subsequent collisions is deterministic and constant, it is possible to obtain a QFI that is orders of magnitude higher than the thermal Fisher information for specific values of the coupling parameters and temperature~\cite{Seah2019,Shu2020}. However, a drawback of this is that such high precision is restricted to a  narrow parameter range, delicately dependent on the temperature of the environment. Such a situation is clearly not ideal given that the temperature is the very quantity which we wish to estimate~\cite{Seah2019,Shu2020}.  Approaches to address this issue include introducing global estimation schemes~\cite{Campbell2017,miller2018energy,mok2021optimal,Mehboudi2021,Jorgensen2021} and biased estimators~\cite{alves2021bayesian}. Here, we demonstrate that if the interactions are random, governed by a particular WTD, this randomicity has an important effect on the value of the parameter $\Delta$ that determines the possible advantage over the thermal Fisher information. It is straightforward to extend the proof of Eq.~\eqref{Eq::Delta} from Ref.~\cite{Shu2020} to random waiting time distributions
\begin{equation}
    \mathcal{F}^{\ket{+_x}}_N = \mathcal{F}^{th} + \sum_{i=1}^{N-1} \Delta_i,
\end{equation}
where $\Delta_i$ takes an identical form as given in Eq.~\eqref{Eq::Delta} except with $\tau_{S E}$ now replaced with a variable time $\tau_{S E}^i$. To determine the average performance of a particular WTD, $p(t)$, we now average over each collision time
\begin{equation}
    \mathscr{F}^{\ket{+_x}}_N = \int_0^\infty\cdots\int_0^\infty\prod_{i=1}^{N-1} d\tau_{S E}^i p(\tau_{S E}^i) \mathcal{F}^{\ket{+_x}}_N = \mathcal{F}^{th} + (N-1)\Bar{\Delta},
\end{equation}
where $\overline{\Delta} \!=\! \int_0^\infty d\tau_{S E}\, p(\tau_{S E})\,\Delta$ with the WTD $p(t)$ being any positive function that satisfies $\int_0^\infty dt\, p(t) \!=\! 1$. In Fig.~\ref{fig:ZZratio} we show the (log of the) ratio between $\overline{\Delta}$ for an exponential distribution, i.e. $k\!=\!1$ (arbitrary choice) and the deterministic $\Delta$. We can see that the randomness allows for a significant performance boost (up to 10 times larger) over a wide range of parameters at the cost of a slight sensitivity loss when the deterministic QFI is maximal.

While we have established that an advantage over the regularly spaced collisions can be achieved for a particular choice of WTD, we now turn our attention to how the particular form of distribution affects the performance. As mentioned previously, varying $k$ in the Weibull distribution, Eq.~\eqref{weibull}, interpolates between distributions with regularly space collisions for $k\!\to\! \infty$ to collisions in batches followed by long pauses as $k\!\to \!0$. We compare the QFI for various values of $k$ at a fixed (arbitrarily chosen) value of temperature, corresponding to $\bar{n}\! =\! 2$, in Fig.~\ref{fig:weibull}. For larger values of $k$, we find the behaviour tends to the deterministic case which is characterised by a QFI with a large peak that is narrow in the parameter range. While the scheme is highly effective, it requires the precise knowledge of the coupling between system and environment. However, for smaller values of $k$, leading to a more random sequence of collisions, we find the range over which an advantage can be demonstrated is significantly broadened, albeit at the expense of reducing the ``maximum" achievable precision. Thus, by introducing stochasticity to the process we are able to alleviate the need for precise knowledge of the optimal system-environment coupling. Interestingly there is a limit to how small $k$ can be and still retain an advantage, with very small $k$ leading to collisions happening so close together that no additional information can be gained.

\subsection{Optimal Measurements}
While the QFI places an asymptotic bound on the accuracy of parameter estimation it does not provide details on precisely what POVM should be implemented in order to saturate the bound. Therefore, identifying the measurements that must be performed on the auxiliary units is important for assessing the implementability of the scheme, something which is particularly relevant for our stochastic collisional approach, in order to assess whether optimal measurements depend on the waiting time between collisions. To find the optimal measurement we need the symmetric logarithmic derivative (SLD) operator $L_a$ for parameter $x_a$. In terms of the eigen-decomposition of $\rho \!=\! \sum_{i}\lambda_i\dyad{\lambda_i}$, the SLD operator is~\cite{Liu2019}
\begin{equation*}
    \bra{\lambda_i}L_a\ket{\lambda_j}= \delta_{ij}\frac{\partial_a\lambda_i}{\lambda_i}+\frac{2(\lambda_j-\lambda_i)}{\lambda_i+\lambda_j}\bra{\lambda_i}\partial_a\lambda_j\rangle.
\end{equation*}
For our scheme we find that the eigenvectors $\{\ket{l_i}\}$ of $L_a$ are independent of $x_a$ and the Fisher information is
\begin{equation*}
\begin{split}
    \mathcal{I}_{aa} &=\sum_i\frac{\bra{l_i}\partial_a\rho\ket{l_i}^2}{\bra{l_i}\rho\ket{l_i}}
    = \sum_i\frac{\bra{l_i}\rho L_a + L_a\rho\ket{l_i}^2}{\bra{l_i}\rho\ket{l_i}}\\ 
    &= \tr{\rho L_a^2} = \mathcal{F}_{aa}
\end{split}
\end{equation*}
where $\mathcal{F}_{aa}$ is the quantum Fisher information for parameter $x_a$. This implies that the optimal measurement corresponds to one performed over the $\{\ket{l_i}\}$ basis. For the $ZZ$ interaction considered \steve{with $g\tau_{SA}\!=\!\pi/2$}, the eigenvectors $\ket{\lambda_i}$ of $\rho$ are independent of $T$ and $\gamma$ meaning that the optimal measurement is precisely the measurement in the $\{\ket{\lambda_i}\}$ basis and is the same for both $T$ and $\gamma$. For the auxiliary units initialised in the $\ket{x_+}$ state considered here, there is some ambiguity in the measurement basis due to degeneracy in the eigenvalues. However, the simplest basis is ${\ket{y_i}\dots\ket{y_j}}$ for $i,j \!\in\! \{+,-\}$ and $y_+ \! =\! \{1,i\}$, $y_- = \{1,-i\}$, with this result holding for any number of auxiliary units. Thus, the optimal measurements involve only product states and therefore can be performed using only local, single-qubit projective measurements, which is consistent with the results of Sec.~\ref{correlations} where we established that there are no genuinely quantum correlations present in the state.

\subsection{Partial Swap Interactions}
\label{sec:Swap}
We conclude our analysis by considering an alternative form for the $S$-$A_i$ interaction that has been considered frequently in collisional thermometry~\cite{Seah2019,Shu2020,alves2021bayesian}. The partial swap (also referred to as an exchange) interaction is given by
\begin{equation}
    H_{SA}^{\text{Swap}} = \hbar g (\sigma^+_S\sigma^-_A+\sigma^-_S\sigma^+_A).
\end{equation}
where similarly to the previous case we are able to tune the effective couplings, $\gamma \tau_{SE}$ and $g \tau_{SA}$. In contrast to the $ZZ$ interaction, now the system and auxiliaries will exchange energy as well as coherences and thus the intermediary system will not remain in the Gibbs state throughout the dynamics. As a consequence it is possible to gain a significant advantage over the thermal Fisher information from just a single auxiliary unit. When a single collision corresponds to a full swap the QFI is maximised and there are no correlations established between subsequent collisional units. Clearly, this interaction is highly disruptive to the intermediary system. We now find that the optimal state for each $A_i$ is the ground state~\cite{Shu2020} and the corresponding optimal measurement being projective measurements in the energy eigenstates. It is worth remarking that a similar advantage could be obtained by measuring the intermediary system directly and not letting it fully thermalise in between measurements~\cite{DePasquale2017}. While clearly there are some differences due to the change in interaction, we find that introducing different waiting time distributions has a qualitatively identical effect in this case, i.e. the introduction of stochasticity allows to significantly extend the range over which a thermometric advantage can be gained from the collisional themometry scheme.

\section{Conclusions}
\label{sec:Concl}
We have extended the framework of collisional quantum thermometry to include stochastic waiting time distributions (WTDs). We demonstrated that introducing a random WTD results in an advantage over the thermal Fisher information for a broader range of parameters, thus alleviating the need to precisely know the coupling strength with the environment. For a dephasing interaction between the collisional units and the intermediary system, we find that only classical correlations between the auxiliary units are established, and that while these correlations appear to be a necessary ingredient to achieve the increased performance, there is not a clear one-to-one relation between attained precision and degree of correlation.

\acknowledgements
We are grateful to Gabriel Landi for insightful discussions. E.O.C and S.C. gratefully acknowledge the Science Foundation Ireland Starting Investigator Research Grant ``SpeedDemon" (No. 18/SIRG/5508) for financial support. B.V. acknowledges the UniMi Transition Grant H2020.

\bibliography{refs}

\begin{thebibliography}{51}%
\makeatletter
\providecommand \@ifxundefined [1]{%
 \@ifx{#1\undefined}
}%
\providecommand \@ifnum [1]{%
 \ifnum #1\expandafter \@firstoftwo
 \else \expandafter \@secondoftwo
 \fi
}%
\providecommand \@ifx [1]{%
 \ifx #1\expandafter \@firstoftwo
 \else \expandafter \@secondoftwo
 \fi
}%
\providecommand \natexlab [1]{#1}%
\providecommand \enquote  [1]{``#1''}%
\providecommand \bibnamefont  [1]{#1}%
\providecommand \bibfnamefont [1]{#1}%
\providecommand \citenamefont [1]{#1}%
\providecommand \href@noop [0]{\@secondoftwo}%
\providecommand \href [0]{\begingroup \@sanitize@url \@href}%
\providecommand \@href[1]{\@@startlink{#1}\@@href}%
\providecommand \@@href[1]{\endgroup#1\@@endlink}%
\providecommand \@sanitize@url [0]{\catcode `\\12\catcode `\$12\catcode
  `\&12\catcode `\#12\catcode `\^12\catcode `\_12\catcode `\%12\relax}%
\providecommand \@@startlink[1]{}%
\providecommand \@@endlink[0]{}%
\providecommand \url  [0]{\begingroup\@sanitize@url \@url }%
\providecommand \@url [1]{\endgroup\@href {#1}{\urlprefix }}%
\providecommand \urlprefix  [0]{URL }%
\providecommand \Eprint [0]{\href }%
\providecommand \doibase [0]{http://dx.doi.org/}%
\providecommand \selectlanguage [0]{\@gobble}%
\providecommand \bibinfo  [0]{\@secondoftwo}%
\providecommand \bibfield  [0]{\@secondoftwo}%
\providecommand \translation [1]{[#1]}%
\providecommand \BibitemOpen [0]{}%
\providecommand \bibitemStop [0]{}%
\providecommand \bibitemNoStop [0]{.\EOS\space}%
\providecommand \EOS [0]{\spacefactor3000\relax}%
\providecommand \BibitemShut  [1]{\csname bibitem#1\endcsname}%
\let\auto@bib@innerbib\@empty
\bibitem [{\citenamefont {Giovannetti}\ \emph {et~al.}(2006)\citenamefont
  {Giovannetti}, \citenamefont {Lloyd},\ and\ \citenamefont
  {Maccone}}]{Giovannetti2006}%
  \BibitemOpen
  \bibfield  {author} {\bibinfo {author} {\bibfnamefont {V.}~\bibnamefont
  {Giovannetti}}, \bibinfo {author} {\bibfnamefont {S.}~\bibnamefont {Lloyd}},
  \ and\ \bibinfo {author} {\bibfnamefont {L.}~\bibnamefont {Maccone}},\
  }\bibfield  {title} {\enquote {\bibinfo {title} {{Quantum metrology}},}\
  }\href {\doibase 10.1103/PhysRevLett.96.010401} {\bibfield  {journal}
  {\bibinfo  {journal} {Phys. Rev. Lett.}\ }\textbf {\bibinfo {volume} {96}},\
  \bibinfo {pages} {010401} (\bibinfo {year} {2006})}\BibitemShut {NoStop}%
\bibitem [{\citenamefont {Ac{\'{i}}n}\ \emph {et~al.}(2018)\citenamefont
  {Ac{\'{i}}n}, \citenamefont {Bloch}, \citenamefont {Buhrman}, \citenamefont
  {Calarco}, \citenamefont {Eichler}, \citenamefont {Eisert}, \citenamefont
  {Esteve}, \citenamefont {Gisin}, \citenamefont {Glaser}, \citenamefont
  {Jelezko}, \citenamefont {Kuhr}, \citenamefont {Lewenstein}, \citenamefont
  {Riedel}, \citenamefont {Schmidt}, \citenamefont {Thew}, \citenamefont
  {Wallraff}, \citenamefont {Walmsley},\ and\ \citenamefont
  {Wilhelm}}]{Acin2018}%
  \BibitemOpen
  \bibfield  {author} {\bibinfo {author} {\bibfnamefont {A.}~\bibnamefont
  {Ac{\'{i}}n}}, \bibinfo {author} {\bibfnamefont {I.}~\bibnamefont {Bloch}},
  \bibinfo {author} {\bibfnamefont {H.}~\bibnamefont {Buhrman}}, \bibinfo
  {author} {\bibfnamefont {T.}~\bibnamefont {Calarco}}, \bibinfo {author}
  {\bibfnamefont {C.}~\bibnamefont {Eichler}}, \bibinfo {author} {\bibfnamefont
  {J.}~\bibnamefont {Eisert}}, \bibinfo {author} {\bibfnamefont
  {D.}~\bibnamefont {Esteve}}, \bibinfo {author} {\bibfnamefont
  {N.}~\bibnamefont {Gisin}}, \bibinfo {author} {\bibfnamefont {S.~J.}\
  \bibnamefont {Glaser}}, \bibinfo {author} {\bibfnamefont {F.}~\bibnamefont
  {Jelezko}}, \bibinfo {author} {\bibfnamefont {S.}~\bibnamefont {Kuhr}},
  \bibinfo {author} {\bibfnamefont {M.}~\bibnamefont {Lewenstein}}, \bibinfo
  {author} {\bibfnamefont {M.~F.}\ \bibnamefont {Riedel}}, \bibinfo {author}
  {\bibfnamefont {P.~O.}\ \bibnamefont {Schmidt}}, \bibinfo {author}
  {\bibfnamefont {R.}~\bibnamefont {Thew}}, \bibinfo {author} {\bibfnamefont
  {A.}~\bibnamefont {Wallraff}}, \bibinfo {author} {\bibfnamefont
  {I.}~\bibnamefont {Walmsley}}, \ and\ \bibinfo {author} {\bibfnamefont
  {F.~K.}\ \bibnamefont {Wilhelm}},\ }\bibfield  {title} {\enquote {\bibinfo
  {title} {{The quantum technologies roadmap: A European community view}},}\
  }\href {\doibase 10.1088/1367-2630/aad1ea} {\bibfield  {journal} {\bibinfo
  {journal} {New J. Phys.}\ }\textbf {\bibinfo {volume} {20}},\ \bibinfo
  {pages} {080201} (\bibinfo {year} {2018})}\BibitemShut {NoStop}%
\bibitem [{\citenamefont {Braun}\ \emph {et~al.}(2018)\citenamefont {Braun},
  \citenamefont {Adesso}, \citenamefont {Benatti}, \citenamefont {Floreanini},
  \citenamefont {Marzolino}, \citenamefont {Mitchell},\ and\ \citenamefont
  {Pirandola}}]{Braun2018}%
  \BibitemOpen
  \bibfield  {author} {\bibinfo {author} {\bibfnamefont {D.}~\bibnamefont
  {Braun}}, \bibinfo {author} {\bibfnamefont {G.}~\bibnamefont {Adesso}},
  \bibinfo {author} {\bibfnamefont {F.}~\bibnamefont {Benatti}}, \bibinfo
  {author} {\bibfnamefont {R.}~\bibnamefont {Floreanini}}, \bibinfo {author}
  {\bibfnamefont {U.}~\bibnamefont {Marzolino}}, \bibinfo {author}
  {\bibfnamefont {M.~W.}\ \bibnamefont {Mitchell}}, \ and\ \bibinfo {author}
  {\bibfnamefont {S.}~\bibnamefont {Pirandola}},\ }\bibfield  {title} {\enquote
  {\bibinfo {title} {{Quantum-enhanced measurements without entanglement}},}\
  }\href {\doibase 10.1103/RevModPhys.90.035006} {\bibfield  {journal}
  {\bibinfo  {journal} {Rev. Mod. Phys.}\ }\textbf {\bibinfo {volume} {90}},\
  \bibinfo {pages} {035006} (\bibinfo {year} {2018})}\BibitemShut {NoStop}%
\bibitem [{\citenamefont {Razavian}\ \emph {et~al.}(2019)\citenamefont
  {Razavian}, \citenamefont {Benedetti}, \citenamefont {Bina}, \citenamefont
  {Akbari-Kourbolagh},\ and\ \citenamefont {Paris}}]{Razavian2019}%
  \BibitemOpen
  \bibfield  {author} {\bibinfo {author} {\bibfnamefont {S.}~\bibnamefont
  {Razavian}}, \bibinfo {author} {\bibfnamefont {C.}~\bibnamefont {Benedetti}},
  \bibinfo {author} {\bibfnamefont {M.}~\bibnamefont {Bina}}, \bibinfo {author}
  {\bibfnamefont {Y.}~\bibnamefont {Akbari-Kourbolagh}}, \ and\ \bibinfo
  {author} {\bibfnamefont {M.~G.A.}\ \bibnamefont {Paris}},\ }\bibfield
  {title} {\enquote {\bibinfo {title} {{Quantum thermometry by single-qubit
  dephasing}},}\ }\href {\doibase 10.1140/epjp/i2019-12708-9} {\bibfield
  {journal} {\bibinfo  {journal} {Eur. Phys. J. Plus}\ }\textbf {\bibinfo
  {volume} {134}},\ \bibinfo {pages} {1--9} (\bibinfo {year}
  {2019})}\BibitemShut {NoStop}%
\bibitem [{\citenamefont {Correa}\ \emph {et~al.}(2015)\citenamefont {Correa},
  \citenamefont {Mehboudi}, \citenamefont {Adesso},\ and\ \citenamefont
  {Sanpera}}]{Correa2015}%
  \BibitemOpen
  \bibfield  {author} {\bibinfo {author} {\bibfnamefont {L.~A.}\ \bibnamefont
  {Correa}}, \bibinfo {author} {\bibfnamefont {M.}~\bibnamefont {Mehboudi}},
  \bibinfo {author} {\bibfnamefont {G.}~\bibnamefont {Adesso}}, \ and\ \bibinfo
  {author} {\bibfnamefont {A.}~\bibnamefont {Sanpera}},\ }\bibfield  {title}
  {\enquote {\bibinfo {title} {{Individual quantum probes for optimal
  thermometry}},}\ }\href {\doibase 10.1103/PhysRevLett.114.220405} {\bibfield
  {journal} {\bibinfo  {journal} {Phys. Rev. Lett.}\ }\textbf {\bibinfo
  {volume} {114}},\ \bibinfo {pages} {220405} (\bibinfo {year}
  {2015})}\BibitemShut {NoStop}%
\bibitem [{\citenamefont {Mitchison}\ \emph {et~al.}(2020)\citenamefont
  {Mitchison}, \citenamefont {Fogarty}, \citenamefont {Guarnieri},
  \citenamefont {Campbell}, \citenamefont {Busch},\ and\ \citenamefont
  {Goold}}]{Mitchison2020}%
  \BibitemOpen
  \bibfield  {author} {\bibinfo {author} {\bibfnamefont {M.~T.}\ \bibnamefont
  {Mitchison}}, \bibinfo {author} {\bibfnamefont {T.}~\bibnamefont {Fogarty}},
  \bibinfo {author} {\bibfnamefont {G.}~\bibnamefont {Guarnieri}}, \bibinfo
  {author} {\bibfnamefont {S.}~\bibnamefont {Campbell}}, \bibinfo {author}
  {\bibfnamefont {Th.}\ \bibnamefont {Busch}}, \ and\ \bibinfo {author}
  {\bibfnamefont {J.}~\bibnamefont {Goold}},\ }\bibfield  {title} {\enquote
  {\bibinfo {title} {{In Situ Thermometry of a Cold Fermi Gas via Dephasing
  Impurities}},}\ }\href {\doibase 10.1103/PhysRevLett.125.080402} {\bibfield
  {journal} {\bibinfo  {journal} {Phys. Rev. Lett.}\ }\textbf {\bibinfo
  {volume} {125}},\ \bibinfo {pages} {080402} (\bibinfo {year}
  {2020})}\BibitemShut {NoStop}%
\bibitem [{\citenamefont {Mehboudi}\ \emph {et~al.}(2019)\citenamefont
  {Mehboudi}, \citenamefont {Sanpera},\ and\ \citenamefont
  {Correa}}]{Mehboudi2019}%
  \BibitemOpen
  \bibfield  {author} {\bibinfo {author} {\bibfnamefont {M.}~\bibnamefont
  {Mehboudi}}, \bibinfo {author} {\bibfnamefont {A.}~\bibnamefont {Sanpera}}, \
  and\ \bibinfo {author} {\bibfnamefont {L.~A.}\ \bibnamefont {Correa}},\
  }\bibfield  {title} {\enquote {\bibinfo {title} {{Thermometry in the quantum
  regime: recent theoretical progress}},}\ }\href {\doibase
  10.1088/1751-8121/AB2828} {\bibfield  {journal} {\bibinfo  {journal} {J.
  Phys. A}\ }\textbf {\bibinfo {volume} {52}},\ \bibinfo {pages} {303001}
  (\bibinfo {year} {2019})}\BibitemShut {NoStop}%
\bibitem [{\citenamefont {Maccone}(2013)}]{Maccone2013}%
  \BibitemOpen
  \bibfield  {author} {\bibinfo {author} {\bibfnamefont {L.}~\bibnamefont
  {Maccone}},\ }\bibfield  {title} {\enquote {\bibinfo {title} {{Intuitive
  reason for the usefulness of entanglement in quantum metrology}},}\ }\href
  {\doibase 10.1103/PhysRevA.88.042109} {\bibfield  {journal} {\bibinfo
  {journal} {Phys. Rev. A}\ }\textbf {\bibinfo {volume} {88}},\ \bibinfo
  {pages} {042109} (\bibinfo {year} {2013})}\BibitemShut {NoStop}%
\bibitem [{\citenamefont {Huang}\ \emph {et~al.}(2016)\citenamefont {Huang},
  \citenamefont {Macchiavello},\ and\ \citenamefont {Maccone}}]{Huang2016}%
  \BibitemOpen
  \bibfield  {author} {\bibinfo {author} {\bibfnamefont {Z.}~\bibnamefont
  {Huang}}, \bibinfo {author} {\bibfnamefont {C.}~\bibnamefont {Macchiavello}},
  \ and\ \bibinfo {author} {\bibfnamefont {L.}~\bibnamefont {Maccone}},\
  }\bibfield  {title} {\enquote {\bibinfo {title} {{Usefulness of
  entanglement-assisted quantum metrology}},}\ }\href {\doibase
  10.1103/PhysRevA.94.012101} {\bibfield  {journal} {\bibinfo  {journal} {Phys.
  Rev. A}\ }\textbf {\bibinfo {volume} {94}},\ \bibinfo {pages} {012101}
  (\bibinfo {year} {2016})}\BibitemShut {NoStop}%
\bibitem [{\citenamefont {Micadei}\ \emph {et~al.}(2015)\citenamefont
  {Micadei}, \citenamefont {Rowlands}, \citenamefont {Pollock}, \citenamefont
  {C{\'{e}}leri}, \citenamefont {Serra},\ and\ \citenamefont
  {Modi}}]{Micadei2015}%
  \BibitemOpen
  \bibfield  {author} {\bibinfo {author} {\bibfnamefont {K.}~\bibnamefont
  {Micadei}}, \bibinfo {author} {\bibfnamefont {D.~A.}\ \bibnamefont
  {Rowlands}}, \bibinfo {author} {\bibfnamefont {F.~A.}\ \bibnamefont
  {Pollock}}, \bibinfo {author} {\bibfnamefont {L.~C.}\ \bibnamefont
  {C{\'{e}}leri}}, \bibinfo {author} {\bibfnamefont {R.~M.}\ \bibnamefont
  {Serra}}, \ and\ \bibinfo {author} {\bibfnamefont {K.}~\bibnamefont {Modi}},\
  }\bibfield  {title} {\enquote {\bibinfo {title} {{Coherent measurements in
  quantum metrology}},}\ }\href {\doibase 10.1088/1367-2630/17/2/023057}
  {\bibfield  {journal} {\bibinfo  {journal} {New J. Phys.}\ }\textbf {\bibinfo
  {volume} {17}},\ \bibinfo {pages} {023057} (\bibinfo {year}
  {2015})}\BibitemShut {NoStop}%
\bibitem [{\citenamefont {Pires}\ \emph {et~al.}(2018)\citenamefont {Pires},
  \citenamefont {Silva}, \citenamefont {Deazevedo}, \citenamefont
  {Soares-Pinto},\ and\ \citenamefont {Filgueiras}}]{Pires2018}%
  \BibitemOpen
  \bibfield  {author} {\bibinfo {author} {\bibfnamefont {D.~P.}\ \bibnamefont
  {Pires}}, \bibinfo {author} {\bibfnamefont {I.~A.}\ \bibnamefont {Silva}},
  \bibinfo {author} {\bibfnamefont {E.~R.}\ \bibnamefont {Deazevedo}}, \bibinfo
  {author} {\bibfnamefont {D.~O.}\ \bibnamefont {Soares-Pinto}}, \ and\
  \bibinfo {author} {\bibfnamefont {J.~G.}\ \bibnamefont {Filgueiras}},\
  }\bibfield  {title} {\enquote {\bibinfo {title} {{Coherence orders,
  decoherence, and quantum metrology}},}\ }\href {\doibase
  10.1103/PhysRevA.98.032101} {\bibfield  {journal} {\bibinfo  {journal} {Phys.
  Rev. A}\ }\textbf {\bibinfo {volume} {98}},\ \bibinfo {pages} {032101}
  (\bibinfo {year} {2018})}\BibitemShut {NoStop}%
\bibitem [{\citenamefont {Castellini}\ \emph {et~al.}(2019)\citenamefont
  {Castellini}, \citenamefont {{Lo Franco}}, \citenamefont {Lami},
  \citenamefont {Winter}, \citenamefont {Adesso},\ and\ \citenamefont
  {Compagno}}]{Castellini2019}%
  \BibitemOpen
  \bibfield  {author} {\bibinfo {author} {\bibfnamefont {A.}~\bibnamefont
  {Castellini}}, \bibinfo {author} {\bibfnamefont {R.}~\bibnamefont {{Lo
  Franco}}}, \bibinfo {author} {\bibfnamefont {L.}~\bibnamefont {Lami}},
  \bibinfo {author} {\bibfnamefont {A.}~\bibnamefont {Winter}}, \bibinfo
  {author} {\bibfnamefont {G.}~\bibnamefont {Adesso}}, \ and\ \bibinfo {author}
  {\bibfnamefont {G.}~\bibnamefont {Compagno}},\ }\bibfield  {title} {\enquote
  {\bibinfo {title} {{Indistinguishability-enabled coherence for quantum
  metrology}},}\ }\href {\doibase 10.1103/PhysRevA.100.012308} {\bibfield
  {journal} {\bibinfo  {journal} {Phys. Rev. A}\ }\textbf {\bibinfo {volume}
  {100}},\ \bibinfo {pages} {012308} (\bibinfo {year} {2019})}\BibitemShut
  {NoStop}%
\bibitem [{\citenamefont {Campbell}\ \emph {et~al.}(2017)\citenamefont
  {Campbell}, \citenamefont {Mehboudi}, \citenamefont {Chiara},\ and\
  \citenamefont {Paternostro}}]{Campbell2017}%
  \BibitemOpen
  \bibfield  {author} {\bibinfo {author} {\bibfnamefont {S.}~\bibnamefont
  {Campbell}}, \bibinfo {author} {\bibfnamefont {M.}~\bibnamefont {Mehboudi}},
  \bibinfo {author} {\bibfnamefont {G.~De}\ \bibnamefont {Chiara}}, \ and\
  \bibinfo {author} {\bibfnamefont {M.}~\bibnamefont {Paternostro}},\
  }\bibfield  {title} {\enquote {\bibinfo {title} {{Global and local
  thermometry schemes in coupled quantum systems}},}\ }\href {\doibase
  10.1088/1367-2630/aa7fac} {\bibfield  {journal} {\bibinfo  {journal} {New J.
  Phys.}\ }\textbf {\bibinfo {volume} {19}},\ \bibinfo {pages} {103003}
  (\bibinfo {year} {2017})}\BibitemShut {NoStop}%
\bibitem [{\citenamefont {Jahromi}(2020)}]{Jahromi2020}%
  \BibitemOpen
  \bibfield  {author} {\bibinfo {author} {\bibfnamefont {H.~R.}\ \bibnamefont
  {Jahromi}},\ }\bibfield  {title} {\enquote {\bibinfo {title} {{Quantum
  thermometry in a squeezed thermal bath}},}\ }\href {\doibase
  10.1088/1402-4896/ab4de5} {\bibfield  {journal} {\bibinfo  {journal} {Physica
  Scripta}\ }\textbf {\bibinfo {volume} {95}},\ \bibinfo {pages} {035107}
  (\bibinfo {year} {2020})}\BibitemShut {NoStop}%
\bibitem [{\citenamefont {Correa}\ \emph {et~al.}(2017)\citenamefont {Correa},
  \citenamefont {Perarnau-Llobet}, \citenamefont {Hovhannisyan}, \citenamefont
  {Hern{\'{a}}ndez-Santana}, \citenamefont {Mehboudi},\ and\ \citenamefont
  {Sanpera}}]{Correa2017}%
  \BibitemOpen
  \bibfield  {author} {\bibinfo {author} {\bibfnamefont {L.~A.}\ \bibnamefont
  {Correa}}, \bibinfo {author} {\bibfnamefont {M.}~\bibnamefont
  {Perarnau-Llobet}}, \bibinfo {author} {\bibfnamefont {K.~V.}\ \bibnamefont
  {Hovhannisyan}}, \bibinfo {author} {\bibfnamefont {S.}~\bibnamefont
  {Hern{\'{a}}ndez-Santana}}, \bibinfo {author} {\bibfnamefont
  {M.}~\bibnamefont {Mehboudi}}, \ and\ \bibinfo {author} {\bibfnamefont
  {A.}~\bibnamefont {Sanpera}},\ }\bibfield  {title} {\enquote {\bibinfo
  {title} {{Enhancement of low temperature thermometry by strong coupling}},}\
  }\href {\doibase 10.1103/PhysRevA.96.062103} {\bibfield  {journal} {\bibinfo
  {journal} {Phys. Rev. A}\ }\textbf {\bibinfo {volume} {96}},\ \bibinfo
  {pages} {062103} (\bibinfo {year} {2017})}\BibitemShut {NoStop}%
\bibitem [{\citenamefont {Hovhannisyan}\ and\ \citenamefont
  {Correa}(2018)}]{Hovhannisyan2018}%
  \BibitemOpen
  \bibfield  {author} {\bibinfo {author} {\bibfnamefont {K.~V.}\ \bibnamefont
  {Hovhannisyan}}\ and\ \bibinfo {author} {\bibfnamefont {L.~A.}\ \bibnamefont
  {Correa}},\ }\bibfield  {title} {\enquote {\bibinfo {title} {{Measuring the
  temperature of cold many-body quantum systems}},}\ }\href {\doibase
  10.1103/PhysRevB.98.045101} {\bibfield  {journal} {\bibinfo  {journal} {Phys.
  Rev. B}\ }\textbf {\bibinfo {volume} {98}},\ \bibinfo {pages} {045101}
  (\bibinfo {year} {2018})}\BibitemShut {NoStop}%
\bibitem [{\citenamefont {Ivanov}(2019)}]{Ivanov2019}%
  \BibitemOpen
  \bibfield  {author} {\bibinfo {author} {\bibfnamefont {P.~A.}\ \bibnamefont
  {Ivanov}},\ }\bibfield  {title} {\enquote {\bibinfo {title} {{Quantum
  thermometry with trapped ions}},}\ }\href {\doibase
  10.1016/j.optcom.2018.12.013} {\bibfield  {journal} {\bibinfo  {journal}
  {Optics Communications}\ }\textbf {\bibinfo {volume} {436}},\ \bibinfo
  {pages} {101--107} (\bibinfo {year} {2019})}\BibitemShut {NoStop}%
\bibitem [{\citenamefont {Jevtic}\ \emph {et~al.}(2015)\citenamefont {Jevtic},
  \citenamefont {Newman}, \citenamefont {Rudolph},\ and\ \citenamefont
  {Stace}}]{Jevtic2015}%
  \BibitemOpen
  \bibfield  {author} {\bibinfo {author} {\bibfnamefont {S.}~\bibnamefont
  {Jevtic}}, \bibinfo {author} {\bibfnamefont {D.}~\bibnamefont {Newman}},
  \bibinfo {author} {\bibfnamefont {T.}~\bibnamefont {Rudolph}}, \ and\
  \bibinfo {author} {\bibfnamefont {T.~M.}\ \bibnamefont {Stace}},\ }\bibfield
  {title} {\enquote {\bibinfo {title} {{Single-qubit thermometry}},}\ }\href
  {\doibase 10.1103/PhysRevA.91.012331} {\bibfield  {journal} {\bibinfo
  {journal} {Phys. Rev. A}\ }\textbf {\bibinfo {volume} {91}},\ \bibinfo
  {pages} {12331} (\bibinfo {year} {2015})}\BibitemShut {NoStop}%
\bibitem [{\citenamefont {Kiilerich}\ \emph {et~al.}(2018)\citenamefont
  {Kiilerich}, \citenamefont {{De Pasquale}},\ and\ \citenamefont
  {Giovannetti}}]{Kiilerich2018}%
  \BibitemOpen
  \bibfield  {author} {\bibinfo {author} {\bibfnamefont {A.~H.}\ \bibnamefont
  {Kiilerich}}, \bibinfo {author} {\bibfnamefont {A.}~\bibnamefont {{De
  Pasquale}}}, \ and\ \bibinfo {author} {\bibfnamefont {V.}~\bibnamefont
  {Giovannetti}},\ }\bibfield  {title} {\enquote {\bibinfo {title} {{Dynamical
  approach to ancilla-assisted quantum thermometry}},}\ }\href {\doibase
  10.1103/PhysRevA.98.042124} {\bibfield  {journal} {\bibinfo  {journal} {Phys.
  Rev. A}\ }\textbf {\bibinfo {volume} {98}},\ \bibinfo {pages} {042124}
  (\bibinfo {year} {2018})}\BibitemShut {NoStop}%
\bibitem [{\citenamefont {Mancino}\ \emph {et~al.}(2017)\citenamefont
  {Mancino}, \citenamefont {Sbroscia}, \citenamefont {Gianani}, \citenamefont
  {Roccia},\ and\ \citenamefont {Barbieri}}]{Mancino2017}%
  \BibitemOpen
  \bibfield  {author} {\bibinfo {author} {\bibfnamefont {L.}~\bibnamefont
  {Mancino}}, \bibinfo {author} {\bibfnamefont {M.}~\bibnamefont {Sbroscia}},
  \bibinfo {author} {\bibfnamefont {I.}~\bibnamefont {Gianani}}, \bibinfo
  {author} {\bibfnamefont {E.}~\bibnamefont {Roccia}}, \ and\ \bibinfo {author}
  {\bibfnamefont {M.}~\bibnamefont {Barbieri}},\ }\bibfield  {title} {\enquote
  {\bibinfo {title} {{Quantum Simulation of Single-Qubit Thermometry Using
  Linear Optics}},}\ }\href {\doibase 10.1103/PhysRevLett.118.130502}
  {\bibfield  {journal} {\bibinfo  {journal} {Phys. Rev. Lett.}\ }\textbf
  {\bibinfo {volume} {118}},\ \bibinfo {pages} {130502} (\bibinfo {year}
  {2017})}\BibitemShut {NoStop}%
\bibitem [{\citenamefont {Potts}\ \emph {et~al.}(2019)\citenamefont {Potts},
  \citenamefont {Brask},\ and\ \citenamefont {Brunner}}]{Potts2019}%
  \BibitemOpen
  \bibfield  {author} {\bibinfo {author} {\bibfnamefont {P.~P.}\ \bibnamefont
  {Potts}}, \bibinfo {author} {\bibfnamefont {J.~B.}\ \bibnamefont {Brask}}, \
  and\ \bibinfo {author} {\bibfnamefont {N.}~\bibnamefont {Brunner}},\
  }\bibfield  {title} {\enquote {\bibinfo {title} {{Fundamental limits on
  low-temperature quantum thermometry with finite resolution}},}\ }\href
  {\doibase 10.22331/q-2019-07-09-161} {\bibfield  {journal} {\bibinfo
  {journal} {Quantum}\ }\textbf {\bibinfo {volume} {3}},\ \bibinfo {pages}
  {161} (\bibinfo {year} {2019})}\BibitemShut {NoStop}%
\bibitem [{\citenamefont {Campbell}\ and\ \citenamefont
  {Vacchini}(2021)}]{Campbell2021a}%
  \BibitemOpen
  \bibfield  {author} {\bibinfo {author} {\bibfnamefont {S.}~\bibnamefont
  {Campbell}}\ and\ \bibinfo {author} {\bibfnamefont {B.}~\bibnamefont
  {Vacchini}},\ }\bibfield  {title} {\enquote {\bibinfo {title} {Collision
  models in open system dynamics: A versatile tool for deeper insights?}}\
  }\href {\doibase 10.1209/0295-5075/133/60001} {\bibfield  {journal} {\bibinfo
   {journal} {{EPL} (Europhysics Letters)}\ }\textbf {\bibinfo {volume}
  {133}},\ \bibinfo {pages} {60001} (\bibinfo {year} {2021})}\BibitemShut
  {NoStop}%
\bibitem [{\citenamefont {Ciccarello}\ \emph {et~al.}(2021)\citenamefont
  {Ciccarello}, \citenamefont {Lorenzo}, \citenamefont {Giovannetti},\ and\
  \citenamefont {Palma}}]{ciccarello2021quantum}%
  \BibitemOpen
  \bibfield  {author} {\bibinfo {author} {\bibfnamefont {F.}~\bibnamefont
  {Ciccarello}}, \bibinfo {author} {\bibfnamefont {S.}~\bibnamefont {Lorenzo}},
  \bibinfo {author} {\bibfnamefont {V.}~\bibnamefont {Giovannetti}}, \ and\
  \bibinfo {author} {\bibfnamefont {G~M.}\ \bibnamefont {Palma}},\ }\bibfield
  {title} {\enquote {\bibinfo {title} {Quantum collision models: open system
  dynamics from repeated interactions},}\ }\href
  {https://arxiv.org/abs/2106.11974} {\bibfield  {journal} {\bibinfo  {journal}
  {arXiv preprint arXiv:2106.11974}\ } (\bibinfo {year} {2021})}\BibitemShut
  {NoStop}%
\bibitem [{\citenamefont {Lorenzo}\ \emph {et~al.}(2017)\citenamefont
  {Lorenzo}, \citenamefont {Ciccarello},\ and\ \citenamefont
  {Palma}}]{Lorenzo2017}%
  \BibitemOpen
  \bibfield  {author} {\bibinfo {author} {\bibfnamefont {S.}~\bibnamefont
  {Lorenzo}}, \bibinfo {author} {\bibfnamefont {F.}~\bibnamefont {Ciccarello}},
  \ and\ \bibinfo {author} {\bibfnamefont {G.~M.}\ \bibnamefont {Palma}},\
  }\bibfield  {title} {\enquote {\bibinfo {title} {{Composite quantum collision
  models}},}\ }\href {\doibase 10.1103/PhysRevA.96.032107} {\bibfield
  {journal} {\bibinfo  {journal} {Phys. Rev. A}\ }\textbf {\bibinfo {volume}
  {96}},\ \bibinfo {pages} {032107} (\bibinfo {year} {2017})}\BibitemShut
  {NoStop}%
\bibitem [{\citenamefont {{De Chiara}}\ and\ \citenamefont
  {Antezza}(2020)}]{DeChiara2020}%
  \BibitemOpen
  \bibfield  {author} {\bibinfo {author} {\bibfnamefont {G.}~\bibnamefont {{De
  Chiara}}}\ and\ \bibinfo {author} {\bibfnamefont {M.}~\bibnamefont
  {Antezza}},\ }\bibfield  {title} {\enquote {\bibinfo {title} {{Quantum
  machines powered by correlated baths}},}\ }\href {\doibase
  10.1103/physrevresearch.2.033315} {\bibfield  {journal} {\bibinfo  {journal}
  {Phys. Rev. Research}\ }\textbf {\bibinfo {volume} {2}},\ \bibinfo {pages}
  {033315} (\bibinfo {year} {2020})}\BibitemShut {NoStop}%
\bibitem [{\citenamefont {Taranto}\ \emph {et~al.}(2020)\citenamefont
  {Taranto}, \citenamefont {Bakhshinezhad}, \citenamefont {Sch{\"{u}}ttelkopf},
  \citenamefont {Clivaz},\ and\ \citenamefont {Huber}}]{Taranto2020}%
  \BibitemOpen
  \bibfield  {author} {\bibinfo {author} {\bibfnamefont {P.}~\bibnamefont
  {Taranto}}, \bibinfo {author} {\bibfnamefont {F.}~\bibnamefont
  {Bakhshinezhad}}, \bibinfo {author} {\bibfnamefont {P.}~\bibnamefont
  {Sch{\"{u}}ttelkopf}}, \bibinfo {author} {\bibfnamefont {F.}~\bibnamefont
  {Clivaz}}, \ and\ \bibinfo {author} {\bibfnamefont {M.}~\bibnamefont
  {Huber}},\ }\bibfield  {title} {\enquote {\bibinfo {title} {{Exponential
  Improvement for Quantum Cooling through Finite-Memory Effects}},}\ }\href
  {\doibase 10.1103/PhysRevApplied.14.054005} {\bibfield  {journal} {\bibinfo
  {journal} {Phys. Rev. Applied}\ }\textbf {\bibinfo {volume} {14}},\ \bibinfo
  {pages} {054005} (\bibinfo {year} {2020})}\BibitemShut {NoStop}%
\bibitem [{\citenamefont {Campbell}\ \emph {et~al.}(2018)\citenamefont
  {Campbell}, \citenamefont {Ciccarello}, \citenamefont {Palma},\ and\
  \citenamefont {Vacchini}}]{Campbell2018}%
  \BibitemOpen
  \bibfield  {author} {\bibinfo {author} {\bibfnamefont {S.}~\bibnamefont
  {Campbell}}, \bibinfo {author} {\bibfnamefont {F.}~\bibnamefont
  {Ciccarello}}, \bibinfo {author} {\bibfnamefont {G.~M.}\ \bibnamefont
  {Palma}}, \ and\ \bibinfo {author} {\bibfnamefont {B.}~\bibnamefont
  {Vacchini}},\ }\bibfield  {title} {\enquote {\bibinfo {title}
  {{System-environment correlations and Markovian embedding of quantum
  non-Markovian dynamics}},}\ }\href {\doibase 10.1103/PhysRevA.98.012142}
  {\bibfield  {journal} {\bibinfo  {journal} {Phys. Rev. A}\ }\textbf {\bibinfo
  {volume} {98}},\ \bibinfo {pages} {012142} (\bibinfo {year}
  {2018})}\BibitemShut {NoStop}%
\bibitem [{\citenamefont {McCloskey}\ and\ \citenamefont
  {Paternostro}(2014)}]{McCloskey2014}%
  \BibitemOpen
  \bibfield  {author} {\bibinfo {author} {\bibfnamefont {R.}~\bibnamefont
  {McCloskey}}\ and\ \bibinfo {author} {\bibfnamefont {M.}~\bibnamefont
  {Paternostro}},\ }\bibfield  {title} {\enquote {\bibinfo {title}
  {{Non-Markovianity and system-environment correlations in a microscopic
  collision model}},}\ }\href {\doibase 10.1103/PhysRevA.89.052120} {\bibfield
  {journal} {\bibinfo  {journal} {Phys. Rev. A}\ }\textbf {\bibinfo {volume}
  {89}},\ \bibinfo {pages} {052120} (\bibinfo {year} {2014})}\BibitemShut
  {NoStop}%
\bibitem [{\citenamefont {{De Chiara}}\ \emph {et~al.}(2018)\citenamefont {{De
  Chiara}}, \citenamefont {Landi}, \citenamefont {Hewgill}, \citenamefont
  {Reid}, \citenamefont {Ferraro}, \citenamefont {Roncaglia},\ and\
  \citenamefont {Antezza}}]{DeChiara2018}%
  \BibitemOpen
  \bibfield  {author} {\bibinfo {author} {\bibfnamefont {G.}~\bibnamefont {{De
  Chiara}}}, \bibinfo {author} {\bibfnamefont {G.}~\bibnamefont {Landi}},
  \bibinfo {author} {\bibfnamefont {A.}~\bibnamefont {Hewgill}}, \bibinfo
  {author} {\bibfnamefont {B.}~\bibnamefont {Reid}}, \bibinfo {author}
  {\bibfnamefont {A.}~\bibnamefont {Ferraro}}, \bibinfo {author} {\bibfnamefont
  {A.~J.}\ \bibnamefont {Roncaglia}}, \ and\ \bibinfo {author} {\bibfnamefont
  {M.}~\bibnamefont {Antezza}},\ }\bibfield  {title} {\enquote {\bibinfo
  {title} {{Reconciliation of quantum local master equations with
  thermodynamics}},}\ }\href {\doibase 10.1088/1367-2630/aaecee} {\bibfield
  {journal} {\bibinfo  {journal} {New J. Phys.}\ }\textbf {\bibinfo {volume}
  {20}},\ \bibinfo {pages} {113024} (\bibinfo {year} {2018})}\BibitemShut
  {NoStop}%
\bibitem [{\citenamefont {Grimmer}\ \emph {et~al.}(2016)\citenamefont
  {Grimmer}, \citenamefont {Layden}, \citenamefont {Mann},\ and\ \citenamefont
  {Mart{\'{i}}n-Mart{\'{i}}nez}}]{Grimmer2016}%
  \BibitemOpen
  \bibfield  {author} {\bibinfo {author} {\bibfnamefont {D.}~\bibnamefont
  {Grimmer}}, \bibinfo {author} {\bibfnamefont {D.}~\bibnamefont {Layden}},
  \bibinfo {author} {\bibfnamefont {R.~B.}\ \bibnamefont {Mann}}, \ and\
  \bibinfo {author} {\bibfnamefont {E.}~\bibnamefont
  {Mart{\'{i}}n-Mart{\'{i}}nez}},\ }\bibfield  {title} {\enquote {\bibinfo
  {title} {{Open dynamics under rapid repeated interaction}},}\ }\href
  {\doibase 10.1103/PhysRevA.94.032126} {\bibfield  {journal} {\bibinfo
  {journal} {Phys. Rev. A}\ }\textbf {\bibinfo {volume} {94}},\ \bibinfo
  {pages} {032126} (\bibinfo {year} {2016})}\BibitemShut {NoStop}%
\bibitem [{\citenamefont {Scarani}\ \emph {et~al.}(2002)\citenamefont
  {Scarani}, \citenamefont {Ziman}, \citenamefont {{\v{S}}telmachovi{\v{c}}},
  \citenamefont {Gisin},\ and\ \citenamefont {Bu{\v{z}}ek}}]{Scarani2002}%
  \BibitemOpen
  \bibfield  {author} {\bibinfo {author} {\bibfnamefont {V.}~\bibnamefont
  {Scarani}}, \bibinfo {author} {\bibfnamefont {M.}~\bibnamefont {Ziman}},
  \bibinfo {author} {\bibfnamefont {P.}~\bibnamefont
  {{\v{S}}telmachovi{\v{c}}}}, \bibinfo {author} {\bibfnamefont
  {N.}~\bibnamefont {Gisin}}, \ and\ \bibinfo {author} {\bibfnamefont
  {V.}~\bibnamefont {Bu{\v{z}}ek}},\ }\bibfield  {title} {\enquote {\bibinfo
  {title} {{Thermalizing quantum machines: Dissipation and entanglement}},}\
  }\href {\doibase 10.1103/PhysRevLett.88.097905} {\bibfield  {journal}
  {\bibinfo  {journal} {Phys. Rev. Lett.}\ }\textbf {\bibinfo {volume} {88}},\
  \bibinfo {pages} {097905} (\bibinfo {year} {2002})}\BibitemShut {NoStop}%
\bibitem [{\citenamefont {Strasberg}\ \emph {et~al.}(2017)\citenamefont
  {Strasberg}, \citenamefont {Schaller}, \citenamefont {Brandes},\ and\
  \citenamefont {Esposito}}]{Strasberg2017}%
  \BibitemOpen
  \bibfield  {author} {\bibinfo {author} {\bibfnamefont {P.}~\bibnamefont
  {Strasberg}}, \bibinfo {author} {\bibfnamefont {G.}~\bibnamefont {Schaller}},
  \bibinfo {author} {\bibfnamefont {T.}~\bibnamefont {Brandes}}, \ and\
  \bibinfo {author} {\bibfnamefont {M.}~\bibnamefont {Esposito}},\ }\bibfield
  {title} {\enquote {\bibinfo {title} {{Quantum and information thermodynamics:
  A unifying framework based on repeated interactions}},}\ }\href {\doibase
  10.1103/PhysRevX.7.021003} {\bibfield  {journal} {\bibinfo  {journal} {Phys.
  Rev. X}\ }\textbf {\bibinfo {volume} {7}},\ \bibinfo {pages} {021003}
  (\bibinfo {year} {2017})}\BibitemShut {NoStop}%
\bibitem [{\citenamefont {Seah}\ \emph {et~al.}(2019)\citenamefont {Seah},
  \citenamefont {Nimmrichter}, \citenamefont {Grimmer}, \citenamefont {Santos},
  \citenamefont {Scarani},\ and\ \citenamefont {Landi}}]{Seah2019}%
  \BibitemOpen
  \bibfield  {author} {\bibinfo {author} {\bibfnamefont {S.}~\bibnamefont
  {Seah}}, \bibinfo {author} {\bibfnamefont {S.}~\bibnamefont {Nimmrichter}},
  \bibinfo {author} {\bibfnamefont {D.}~\bibnamefont {Grimmer}}, \bibinfo
  {author} {\bibfnamefont {J.~P.}\ \bibnamefont {Santos}}, \bibinfo {author}
  {\bibfnamefont {V.}~\bibnamefont {Scarani}}, \ and\ \bibinfo {author}
  {\bibfnamefont {G.~T.}\ \bibnamefont {Landi}},\ }\bibfield  {title} {\enquote
  {\bibinfo {title} {{Collisional Quantum Thermometry}},}\ }\href {\doibase
  10.1103/PhysRevLett.123.180602} {\bibfield  {journal} {\bibinfo  {journal}
  {Phys. Rev. Lett.}\ }\textbf {\bibinfo {volume} {123}},\ \bibinfo {pages}
  {180602} (\bibinfo {year} {2019})}\BibitemShut {NoStop}%
\bibitem [{\citenamefont {Shu}\ \emph {et~al.}(2020)\citenamefont {Shu},
  \citenamefont {Seah},\ and\ \citenamefont {Scarani}}]{Shu2020}%
  \BibitemOpen
  \bibfield  {author} {\bibinfo {author} {\bibfnamefont {A.}~\bibnamefont
  {Shu}}, \bibinfo {author} {\bibfnamefont {S.}~\bibnamefont {Seah}}, \ and\
  \bibinfo {author} {\bibfnamefont {V.}~\bibnamefont {Scarani}},\ }\bibfield
  {title} {\enquote {\bibinfo {title} {{Surpassing the thermal Cram{\'{e}}r-Rao
  bound with collisional thermometry}},}\ }\href {\doibase
  10.1103/PhysRevA.102.042417} {\bibfield  {journal} {\bibinfo  {journal}
  {Phys. Rev. A}\ }\textbf {\bibinfo {volume} {102}},\ \bibinfo {pages}
  {042417} (\bibinfo {year} {2020})}\BibitemShut {NoStop}%
\bibitem [{\citenamefont {Alves}\ and\ \citenamefont
  {Landi}(2021)}]{alves2021bayesian}%
  \BibitemOpen
  \bibfield  {author} {\bibinfo {author} {\bibfnamefont {G.~O.}\ \bibnamefont
  {Alves}}\ and\ \bibinfo {author} {\bibfnamefont {G.~T.}\ \bibnamefont
  {Landi}},\ }\bibfield  {title} {\enquote {\bibinfo {title} {Bayesian
  estimation for collisional thermometry},}\ }\href
  {https://arxiv.org/abs/2106.12072} {\bibfield  {journal} {\bibinfo  {journal}
  {arXiv preprint arXiv:2106.12072}\ } (\bibinfo {year} {2021})}\BibitemShut
  {NoStop}%
\bibitem [{\citenamefont {Ollivier}\ and\ \citenamefont
  {Zurek}(2001)}]{Ollivier2001a}%
  \BibitemOpen
  \bibfield  {author} {\bibinfo {author} {\bibfnamefont {H.}~\bibnamefont
  {Ollivier}}\ and\ \bibinfo {author} {\bibfnamefont {W.~H.}\ \bibnamefont
  {Zurek}},\ }\bibfield  {title} {\enquote {\bibinfo {title} {Quantum discord:
  a measure of the quantumness of correlations},}\ }\href {\doibase
  10.1103/PhysRevLett.88.017901} {\bibfield  {journal} {\bibinfo  {journal}
  {Phys. Rev. Lett.}\ }\textbf {\bibinfo {volume} {88}},\ \bibinfo {pages}
  {017901} (\bibinfo {year} {2001})}\BibitemShut {NoStop}%
\bibitem [{\citenamefont {Henderson}\ and\ \citenamefont
  {Vedral}(2001)}]{Henderson2001a}%
  \BibitemOpen
  \bibfield  {author} {\bibinfo {author} {\bibfnamefont {L}~\bibnamefont
  {Henderson}}\ and\ \bibinfo {author} {\bibfnamefont {V}~\bibnamefont
  {Vedral}},\ }\bibfield  {title} {\enquote {\bibinfo {title} {Classical,
  quantum and total correlations},}\ }\href
  {http://stacks.iop.org/0305-4470/34/i=35/a=315} {\bibfield  {journal}
  {\bibinfo  {journal} {J. Phys. A}\ }\textbf {\bibinfo {volume} {34}},\
  \bibinfo {pages} {6899} (\bibinfo {year} {2001})}\BibitemShut {NoStop}%
\bibitem [{\citenamefont {Luiz}\ \emph {et~al.}(2021)\citenamefont {Luiz},
  \citenamefont {Junior}, \citenamefont {Fanchini},\ and\ \citenamefont
  {Landi}}]{luiz2021machine}%
  \BibitemOpen
  \bibfield  {author} {\bibinfo {author} {\bibfnamefont {F.~S.}\ \bibnamefont
  {Luiz}}, \bibinfo {author} {\bibfnamefont {A.}~\bibnamefont {Junior}},
  \bibinfo {author} {\bibfnamefont {F.~F.}\ \bibnamefont {Fanchini}}, \ and\
  \bibinfo {author} {\bibfnamefont {G.~T}\ \bibnamefont {Landi}},\ }\bibfield
  {title} {\enquote {\bibinfo {title} {Machine classification for probe based
  quantum thermometry},}\ }\href {https://arxiv.org/abs/2107.04555} {\bibfield
  {journal} {\bibinfo  {journal} {arXiv preprint arXiv:2107.04555}\ } (\bibinfo
  {year} {2021})}\BibitemShut {NoStop}%
\bibitem [{\citenamefont {Liu}\ \emph {et~al.}(2019)\citenamefont {Liu},
  \citenamefont {Yuan}, \citenamefont {Lu},\ and\ \citenamefont
  {Wang}}]{Liu2019}%
  \BibitemOpen
  \bibfield  {author} {\bibinfo {author} {\bibfnamefont {J.}~\bibnamefont
  {Liu}}, \bibinfo {author} {\bibfnamefont {H.}~\bibnamefont {Yuan}}, \bibinfo
  {author} {\bibfnamefont {X.-M.}\ \bibnamefont {Lu}}, \ and\ \bibinfo {author}
  {\bibfnamefont {X.}~\bibnamefont {Wang}},\ }\bibfield  {title} {\enquote
  {\bibinfo {title} {{Quantum Fisher information matrix and multiparameter
  estimation}},}\ }\href {\doibase 10.1088/1751-8121/AB5D4D} {\bibfield
  {journal} {\bibinfo  {journal} {J. Phys. A}\ }\textbf {\bibinfo {volume}
  {53}},\ \bibinfo {pages} {023001} (\bibinfo {year} {2019})}\BibitemShut
  {NoStop}%
\bibitem [{\citenamefont {Strasberg}(2019)}]{Strasberg2019}%
  \BibitemOpen
  \bibfield  {author} {\bibinfo {author} {\bibfnamefont {P.}~\bibnamefont
  {Strasberg}},\ }\bibfield  {title} {\enquote {\bibinfo {title} {{Operational
  approach to quantum stochastic thermodynamics}},}\ }\href {\doibase
  10.1103/PhysRevE.100.022127} {\bibfield  {journal} {\bibinfo  {journal}
  {Phys. Rev. E}\ }\textbf {\bibinfo {volume} {100}},\ \bibinfo {pages}
  {022127} (\bibinfo {year} {2019})}\BibitemShut {NoStop}%
\bibitem [{\citenamefont {Vacchini}(2020)}]{Vacchini2020}%
  \BibitemOpen
  \bibfield  {author} {\bibinfo {author} {\bibfnamefont {B.}~\bibnamefont
  {Vacchini}},\ }\bibfield  {title} {\enquote {\bibinfo {title} {{Quantum
  renewal processes}},}\ }\href {\doibase 10.1038/s41598-020-62260-z}
  {\bibfield  {journal} {\bibinfo  {journal} {Sci. Rep.}\ }\textbf {\bibinfo
  {volume} {10}},\ \bibinfo {pages} {1--13} (\bibinfo {year}
  {2020})}\BibitemShut {NoStop}%
\bibitem [{\citenamefont {Chisholm}\ \emph {et~al.}(2021)\citenamefont
  {Chisholm}, \citenamefont {Garc{\'{i}}a-P{\'{e}}rez}, \citenamefont {Rossi},
  \citenamefont {Palma},\ and\ \citenamefont {Maniscalco}}]{Chisholm2021}%
  \BibitemOpen
  \bibfield  {author} {\bibinfo {author} {\bibfnamefont {D.~A}\ \bibnamefont
  {Chisholm}}, \bibinfo {author} {\bibfnamefont {G.}~\bibnamefont
  {Garc{\'{i}}a-P{\'{e}}rez}}, \bibinfo {author} {\bibfnamefont {M.~A.~C.}\
  \bibnamefont {Rossi}}, \bibinfo {author} {\bibfnamefont {G.~M.}\ \bibnamefont
  {Palma}}, \ and\ \bibinfo {author} {\bibfnamefont {S.}~\bibnamefont
  {Maniscalco}},\ }\bibfield  {title} {\enquote {\bibinfo {title} {{Stochastic
  collision model approach to transport phenomena in quantum networks}},}\
  }\href {\doibase 10.1088/1367-2630/ABD57D} {\bibfield  {journal} {\bibinfo
  {journal} {New J. Phys.}\ }\textbf {\bibinfo {volume} {23}},\ \bibinfo
  {pages} {033031} (\bibinfo {year} {2021})}\BibitemShut {NoStop}%
\bibitem [{\citenamefont {Tabanera}\ \emph {et~al.}(2021)\citenamefont
  {Tabanera}, \citenamefont {Luque}, \citenamefont {Jacob}, \citenamefont
  {Esposito}, \citenamefont {Barra},\ and\ \citenamefont
  {Parrondo}}]{Tabanera2021}%
  \BibitemOpen
  \bibfield  {author} {\bibinfo {author} {\bibfnamefont {J.}~\bibnamefont
  {Tabanera}}, \bibinfo {author} {\bibfnamefont {I.}~\bibnamefont {Luque}},
  \bibinfo {author} {\bibfnamefont {S.~L.}\ \bibnamefont {Jacob}}, \bibinfo
  {author} {\bibfnamefont {M.}~\bibnamefont {Esposito}}, \bibinfo {author}
  {\bibfnamefont {F.}~\bibnamefont {Barra}}, \ and\ \bibinfo {author}
  {\bibfnamefont {J.~M.~R.}\ \bibnamefont {Parrondo}},\ }\bibfield  {title}
  {\enquote {\bibinfo {title} {{Quantum collisional thermostats}},}\ }\href
  {https://arxiv.org/abs/2109.10620v1} {\  (\bibinfo {year} {2021})},\ \Eprint
  {http://arxiv.org/abs/2109.10620} {arXiv:2109.10620} \BibitemShut {NoStop}%
\bibitem [{\citenamefont {Ehrich}\ \emph {et~al.}(2020)\citenamefont {Ehrich},
  \citenamefont {Esposito}, \citenamefont {Barra},\ and\ \citenamefont
  {Parrondo}}]{Ehrich2020}%
  \BibitemOpen
  \bibfield  {author} {\bibinfo {author} {\bibfnamefont {J.}~\bibnamefont
  {Ehrich}}, \bibinfo {author} {\bibfnamefont {M.}~\bibnamefont {Esposito}},
  \bibinfo {author} {\bibfnamefont {F.}~\bibnamefont {Barra}}, \ and\ \bibinfo
  {author} {\bibfnamefont {J.~M.~R.}\ \bibnamefont {Parrondo}},\ }\bibfield
  {title} {\enquote {\bibinfo {title} {{Micro-reversibility and thermalization
  with collisional baths}},}\ }\href {\doibase 10.1016/J.PHYSA.2019.122108}
  {\bibfield  {journal} {\bibinfo  {journal} {Physica A}\ }\textbf {\bibinfo
  {volume} {552}},\ \bibinfo {pages} {122108} (\bibinfo {year}
  {2020})}\BibitemShut {NoStop}%
\bibitem [{\citenamefont {Jacob}\ \emph {et~al.}(2021)\citenamefont {Jacob},
  \citenamefont {Esposito}, \citenamefont {Parrondo},\ and\ \citenamefont
  {Barra}}]{Jacob2021}%
  \BibitemOpen
  \bibfield  {author} {\bibinfo {author} {\bibfnamefont {S.~L.}\ \bibnamefont
  {Jacob}}, \bibinfo {author} {\bibfnamefont {M.}~\bibnamefont {Esposito}},
  \bibinfo {author} {\bibfnamefont {J.~M.~R.}\ \bibnamefont {Parrondo}}, \ and\
  \bibinfo {author} {\bibfnamefont {F.}~\bibnamefont {Barra}},\ }\bibfield
  {title} {\enquote {\bibinfo {title} {{Thermalization Induced by Quantum
  Scattering}},}\ }\href {\doibase 10.1103/PRXQuantum.2.020312} {\bibfield
  {journal} {\bibinfo  {journal} {PRX Quantum}\ }\textbf {\bibinfo {volume}
  {2}},\ \bibinfo {pages} {020312} (\bibinfo {year} {2021})}\BibitemShut
  {NoStop}%
\bibitem [{Note1()}]{Note1}%
  \BibitemOpen
  \bibinfo {note} {We remark that our results remain qualitatively unaffected
  for other families of WTD, e.g. Erlang distributions.}\BibitemShut {Stop}%
\bibitem [{\citenamefont {Miller}\ and\ \citenamefont
  {Anders}(2018)}]{miller2018energy}%
  \BibitemOpen
  \bibfield  {author} {\bibinfo {author} {\bibfnamefont {H.~J.~D.}\
  \bibnamefont {Miller}}\ and\ \bibinfo {author} {\bibfnamefont
  {J.}~\bibnamefont {Anders}},\ }\bibfield  {title} {\enquote {\bibinfo {title}
  {Energy-temperature uncertainty relation in quantum thermodynamics},}\ }\href
  {\doibase 10.1038/s41467-018-04536-7} {\bibfield  {journal} {\bibinfo
  {journal} {Nat. Commun.}\ }\textbf {\bibinfo {volume} {9}},\ \bibinfo {pages}
  {1--8} (\bibinfo {year} {2018})}\BibitemShut {NoStop}%
\bibitem [{\citenamefont {Mok}\ \emph {et~al.}(2021)\citenamefont {Mok},
  \citenamefont {Bharti}, \citenamefont {Kwek},\ and\ \citenamefont
  {Bayat}}]{mok2021optimal}%
  \BibitemOpen
  \bibfield  {author} {\bibinfo {author} {\bibfnamefont {W.-K.}\ \bibnamefont
  {Mok}}, \bibinfo {author} {\bibfnamefont {K.}~\bibnamefont {Bharti}},
  \bibinfo {author} {\bibfnamefont {L.-C.}\ \bibnamefont {Kwek}}, \ and\
  \bibinfo {author} {\bibfnamefont {A.}~\bibnamefont {Bayat}},\ }\bibfield
  {title} {\enquote {\bibinfo {title} {Optimal probes for global quantum
  thermometry},}\ }\href {\doibase 10.1038/s42005-021-00572-w} {\bibfield
  {journal} {\bibinfo  {journal} {Communications Physics}\ }\textbf {\bibinfo
  {volume} {4}},\ \bibinfo {pages} {1--8} (\bibinfo {year} {2021})}\BibitemShut
  {NoStop}%
\bibitem [{\citenamefont {Mehboudi}\ \emph {et~al.}(2021)\citenamefont
  {Mehboudi}, \citenamefont {J{\o}rgensen}, \citenamefont {Seah}, \citenamefont
  {Brask}, \citenamefont {Ko{\l}ody{\'{n}}ski},\ and\ \citenamefont
  {Perarnau-Llobet}}]{Mehboudi2021}%
  \BibitemOpen
  \bibfield  {author} {\bibinfo {author} {\bibfnamefont {M.}~\bibnamefont
  {Mehboudi}}, \bibinfo {author} {\bibfnamefont {M.~R.}\ \bibnamefont
  {J{\o}rgensen}}, \bibinfo {author} {\bibfnamefont {S.}~\bibnamefont {Seah}},
  \bibinfo {author} {\bibfnamefont {J.~B.}\ \bibnamefont {Brask}}, \bibinfo
  {author} {\bibfnamefont {J.}~\bibnamefont {Ko{\l}ody{\'{n}}ski}}, \ and\
  \bibinfo {author} {\bibfnamefont {M.}~\bibnamefont {Perarnau-Llobet}},\
  }\bibfield  {title} {\enquote {\bibinfo {title} {{Fundamental limits in
  Bayesian thermometry and attainability via adaptive strategies}},}\ }\href
  {https://arxiv.org/abs/2108.05932v1} {\  (\bibinfo {year} {2021})},\ \Eprint
  {http://arxiv.org/abs/2108.05932} {arXiv:2108.05932} \BibitemShut {NoStop}%
\bibitem [{\citenamefont {J{\o}rgensen}\ \emph {et~al.}(2021)\citenamefont
  {J{\o}rgensen}, \citenamefont {Ko{\l}ody{\'{n}}ski}, \citenamefont
  {Mehboudi}, \citenamefont {Perarnau-Llobet},\ and\ \citenamefont
  {Brask}}]{Jorgensen2021}%
  \BibitemOpen
  \bibfield  {author} {\bibinfo {author} {\bibfnamefont {M.~R.}\ \bibnamefont
  {J{\o}rgensen}}, \bibinfo {author} {\bibfnamefont {J.}~\bibnamefont
  {Ko{\l}ody{\'{n}}ski}}, \bibinfo {author} {\bibfnamefont {M.}~\bibnamefont
  {Mehboudi}}, \bibinfo {author} {\bibfnamefont {M.}~\bibnamefont
  {Perarnau-Llobet}}, \ and\ \bibinfo {author} {\bibfnamefont {J.~B.}\
  \bibnamefont {Brask}},\ }\bibfield  {title} {\enquote {\bibinfo {title}
  {{Bayesian quantum thermometry based on thermodynamic length}},}\ }\href
  {https://arxiv.org/abs/2108.05901v1} {\  (\bibinfo {year} {2021})},\ \Eprint
  {http://arxiv.org/abs/2108.05901} {arXiv:2108.05901} \BibitemShut {NoStop}%
\bibitem [{\citenamefont {{De Pasquale}}\ \emph {et~al.}(2017)\citenamefont
  {{De Pasquale}}, \citenamefont {Yuasa},\ and\ \citenamefont
  {Giovannetti}}]{DePasquale2017}%
  \BibitemOpen
  \bibfield  {author} {\bibinfo {author} {\bibfnamefont {A.}~\bibnamefont {{De
  Pasquale}}}, \bibinfo {author} {\bibfnamefont {K.}~\bibnamefont {Yuasa}}, \
  and\ \bibinfo {author} {\bibfnamefont {V.}~\bibnamefont {Giovannetti}},\
  }\bibfield  {title} {\enquote {\bibinfo {title} {{Estimating temperature via
  sequential measurements}},}\ }\href {\doibase 10.1103/PhysRevA.96.012316}
  {\bibfield  {journal} {\bibinfo  {journal} {Phys. Rev. A}\ }\textbf {\bibinfo
  {volume} {96}},\ \bibinfo {pages} {012316} (\bibinfo {year}
  {2017})}\BibitemShut {NoStop}%
\end{thebibliography}%

\end{document}